\documentclass[a4paper,fleqn,usenatbib]{mnras}

\usepackage{graphicx}	
\usepackage[autostyle]{csquotes}
\usepackage{changepage}
\usepackage{amsmath}	
\usepackage{amssymb}	
\usepackage{epstopdf}

\title[Evolution of Accretion Dynamics of GX 339-4]{Study of Long Term Evolution of Accretion Dynamics of GX 339-4}
\author[Aneesha et al.]{U. Aneesha$^{1}$\thanks{aneesha.14@res.iist.ac.in}, S. Mandal$^{1}$ and H. Sreehari$^{2,3}$\\
$^{1}$Indian Institute of Space Science and Technology (IIST), Trivandrum 695547, India\\
$^{2}$Space Astronomy Group, ISITE Campus, URSC, Outer Ring Road,
Marathahalli, Bangalore, 560037, India\\
$^{3}$Indian Institute of Science, Bangalore, 560012, India}

\date{Accepted XXX. Received YYY; in original form ZZZ}

\pubyear{2019}

\newcommand{\RN}[1]{%
  \textup{\uppercase\expandafter{\romannumeral#1}}%
}
\begin{document}
\label{firstpage}
\maketitle
\begin{abstract}
We study the dynamical behaviour of the galactic black hole source GX 339-4 during 2002-2011 outbursts using RXTE, Swift(XRT), XMM-Newton(PN) archival data. We present the spectral evolution of the source using four outbursts data and discuss their similarities/differences between outbursts. We infer that the second peak in 2002/03 and 2004/05 outbursts can be due to a second instant of triggered instability in the accretion disc due to irradiation from the central X-ray source after peak-I. This propagates in viscous time scale and takes $\sim 80-90$ days after peak-I to produce peak-II. This unifies all four outbursts having a long rising time of $\sim 90$ days. The dynamical evolution of accretion parameters have been studied by modeling the individual observed spectrum with two-component accretion disc model where a Keplerian accretion disc produces the soft photons and the hard part of the spectrum  originates from a hot sub-Keplerian central corona. A generic mathematical model has been proposed to understand the evolution of accretion parameters for sources like GX 339-4 which have longer rising time. Also, the possible differences of physical scenario for outbursts with shorter rising time are also discussed.
\end{abstract}
\begin{keywords}
accretion, accretion discs -- X-rays: binaries -- radiation: dynamics -- individual: GX 339-4
\end{keywords}

\section{Introduction}
Studies have been made over past few decades to understand the accretion physics around black hole X-ray binaries (XRBs). The galactic black hole (BH) candidates in transient low mass X-ray binaries are the ideal systems for the study of accretion physics. Transient XRBs are either in outburst phase with strong X-ray emission ($10^{37}-10^{39}$ erg s$^{-1}$) or in faint ($10^{30}-10^{33}$ erg s$^{-1}$) quiescent phase . 
 
The energy spectra of black hole XRBs are composite in nature: a thermal soft component proposed to have a Keplerian accretion disc \citep{1973A&A....24..337S,1973blho.conf..343N} origin along with a non-thermal hard component. It is generally accepted that the inverse Comptonization \citep[and references therein]{1980A&A....86..121S,1985A&A...143..374S,1994ApJ...434..570T,1998MNRAS.301..435Z} of soft photons from the accretion disc by the hot electron cloud \citep{1995ApJ...452..710N,1995ApJ...455..623C} produces the hard component. Whereas \cite{2005ApJ...635.1203M} suggested non-thermal synchrotron photons as well as thermal and non-thermal synchrotron self-Comptonization in the jet as the source of hard photons. Alternately, inverse Comptonization of photons from the companion star and/or the accretion disc by the base of the jet \citep{2002A&A...388L..25G} can also be the sources of hard photons. The relative contributions of two spectral components and the time variabilities define each spectral state \citep{2005Ap&SS.300..107H,2006ARA&A..44...49R}. In the low/hard state (LHS), the source is faint, the radiation spectrum has a photon index in the range of $1.5 - 1.7$ and X-ray timing variability is maximum \citep{2001PhDT.......226H}. In this state power density spectrum (PDS) is dominated by strong band-limited noise with a strength of 20-50\% rms and type-C quasi periodic oscillation (QPO) is also present. Whereas in high/soft state (HSS) the source is bright, the radiation spectrum has a photon index $\sim 3$, and X-ray timing variability is minimum. The PDS in this state has a power law signature with less than 2-3\% rms value. The state transitions from LHS to HSS pass through the intermediate states (hard intermediate state: HIMS and soft intermediate state:SIMS) where the contribution from accretion disc becomes significant, and also has photon index $\sim 1.7 - 2.5$. The noise in the PDS is weaker (typically 5-20\% rms). Generally, HIMS shows type-C QPOs and SIMS shows type-A and type-B QPOs \citep{2005A&A...440..207B}. The state transition via intermediate state has very short duration ranging from hours to days and hence the accretion physics related to these states is difficult to study.

The global evolution of spectral states during outburst can be understood from different branches of the hardness-intensity diagram (HID) \citep{2005Ap&SS.300..107H, 2006csxs.book..157M, 2006ARA&A..44...49R, 2010LNP...794...53B,2012A&A...542A..56N,2014AdSpR..54.1678R}. The outburst profile starts with a low luminosity having higher hardness and reaches the peak luminosity with a lower hardness. Finally, it comes back again to low luminosity and high hardness at the end of the outburst. This hysteresis nature of HID was first reported by \cite{1995ApJ...442L..13M}.

The source GX 339-4 was discovered in 1973 by satellite OSO-7 \citep{1973ApJ...184L..67M}. The absorption lines in NIR \citep{2017ApJ...846..132H} reveals a giant K-type companion and this confirms the earlier finding \citep{2004ApJ...609..317H, 2008MNRAS.385.2205M}. 
The distance to GX 339-4 is found to be $6 \, \text{kpc}\leq d \leq 15 \, \text{kpc}$ \citep{2004ApJ...609..317H} from studies based on Na D lines whereas \cite{2004MNRAS.351..791Z} has estimated $d \gtrapprox 7 \, \text{kpc}$ from optical and infrared observations. Also, \cite{2016ApJ...821L...6P} have fitted the X-ray spectra with $d=8.4 \pm 0.9 \, \text{kpc}$. Since the system did not show any eclipse in X-ray or optical data, \cite{2002AJ....123.1741C} restricted the upper limit of the inclination angle $i\le 60^\circ$ and \cite{2004MNRAS.351..791Z} suggested a lower limit of $i \geq 45^\circ$ from the secondary mass function estimation. From the reflection modelling of X-ray spectra \cite{2015ApJ...808..122F,2015ApJ...813...84G} have found inclination angle in the range $40^\circ - 60^\circ$ and \cite{2016ApJ...821L...6P} have estimated $i=30^\circ \pm 1^\circ$. Whereas \cite{2017ApJ...846..132H} have quoted $37^\circ < i <78^\circ$ from the NIR absorption lines of the donor star. The optical spectroscopy of Bowenblend and HeII provide a mass function of the system to be $f(M)=5.8 \pm 0.5  \, \textup{M}_{\odot}$ with a binary period of 1.7557 days \citep{2003ApJ...583L..95H}. \cite{2016ApJ...821L...6P} have constrained the mass to be $9.0^{+1.6}_{-1.2} \,  \, \textup{M}_{\odot}$ whereas \cite{2017ApJ...846..132H} have quoted $2.3 \, \textup{M}_{\odot} < \text{M} < 9.5 \, \textup{M}_{\odot}$. Recently, \cite{2019AdSpR..63.1374S} have estimated the mass of the source in the range $8.28-11.89 \, \textup{M}_{\odot}$ from spectral modelling using two-component flow, evolution of QPO frequency and saturation of spectral index. Modelling the reflection component of the X-ray spectra, spin of the system is found to be $a=0.93 \pm 0.01$ \citep{2008ApJ...679L.113M} whereas \cite{2015ApJ...806..262L} have estimated $a>0.97$ and \cite{2016ApJ...821L...6P} have determined $a=0.95^{+0.02}_{-0.08}$ using relxill model for reflection fitting. 

Over past 44 years, the source has shown many successful outbursts as well as many failed outbursts where the source always remain in the LHS during outburst. There are several early reports of state transitions \citep{1997ApJ...479..926M,1999ApJ...519L.159B} and time variability \citep{1984ApJ...285..712M,1991ApJ...383..784M, 2000MNRAS.318..361N} of the source. Detailed timing and spectral studies \citep{2005A&A...440..207B,2010A&A...520A..98D,2012A&A...542A..56N} show a complete evolution of spectral state during outburst. 
Several attempts have been made to understand the spectral behaviour of the source using physically motivated models. \cite{2003A&A...400.1007C}, \cite{2019MNRAS.tmp..596C} used synchrotron emission from the jet and the inverse Comptonization of both the thermal disc photons as well as synchrotron photons to explain radio, IR and X-ray emissions in GX 339-4. \cite{2004MNRAS.351..791Z} has developed an accretion flow model comprises of a cold optically thick outer disc and a hot inner optically thin disc. The RXTE-ASM data during 1998/99 and 2002/03 outburst of GX 339-4 is studied with this model. \cite{2018A&A...617A..46M} have used a unified accretion-ejection model to understand the X-ray and Radio emissions of 2010/11 outburst of GX 339-4. But none of these models address the dynamics of the system self-consistently i.e., given the flow parameters like, accretion rate, angular momentum etc, the flow hydrodynamics should decide the rest. Also sometimes models are not fully self-consistent. For example, in unified accretion-ejection model \citep{2018A&A...617A..46M}, the transition radius between accretion disc and central corona is a parameter and it is independent of the accretion rate which may not be physically correct. Also if the hot central corona is created from the accretion disc, it would not be able to explain the soft time lag \citep{2001ApJ...554L..41S, 2002ApJ...569..362S, 2007ApJ...669.1138S}. Also, not much attention has been paid to understand the accretion dynamics quantitatively by physically modelling the observed spectrum or outburst profiles. Though, \cite{2010ApJ...710L.147M} have qualitatively modelled the general profile HID of the source GRO J1655-40 but without modelling the observed radiation spectra. 

In the present study, we try to understand the spectral evolution as well as evolution of accretion disc parameters from spectral modeling the outbursts of GX 339-4 during 2002-2011 mostly using RXTE data. We have modeled the RXTE (PCA) spectra using both phenomenological model as well as two component accretion flow model \citep{1995ApJ...455..623C,2006ApJ...642L..49C}. As RXTE (PCA) mostly works beyond 3 KeV, we have used XMM-Newton and Swift (XRT) archival data (whenever available) to constrain the soft X-ray part of the energy spectrum. Also, we have done broad band spectral modeling in the energy range 0.3-100.0 keV using XRT/RXTE and XMM-Newton/RXTE data to constrain the high energy spectral index as well as validate the spectral modeling. The outburst profiles differ between outbursts even for the same source and we have tried to understand differences in physical picture  under irradiated disc instability model scenario) between outbursts by studying the evolution of the accretion parameters.
In \S\ref{obs_data}, we discuss the observations and data reduction methods of different instruments used in this study and \S\ref{results}, we present the phenomenological spectral studies and two-component flow model parameters evolution. Finally, in \S\ref{conclusions} we conclude.

\section{Observations and Data Reductions}
\label{obs_data}
In order to understand the evolution of accretion dynamics of GX 339-4 during complete outbursts, all observations between years 2002-2011 using the RXTE, Swift (XRT) and XMM-Newton are analysed and modelled. During this period four successful outbursts of the source have been observed. 

\subsection{RXTE Observations}
We have used standard-2 PCA data (FS4a*.gz) for the spectral analysis. For consistency, the data only from PCU2 are used for analysis as it was always switched on during these period. We have analysed RXTE Proportional Counter Array (PCA) data using {\it HEASOFT software package version 6.14}. We follow the standard FTOOLS task to generate good time interval setting the screening criteria on elevation angle, offset and South Atlantic anomaly (SAA). We generate the background data by comparing the latest bright background model. Source spectrum, background spectrum and response are extracted again following standard FTOOLS tasks. We have considered spectrum in the energy range  3.0 - 60.0 keV without any grouping or binning. A systematic error of 0.5\% is added to the spectrum in 2006/07 and 2010/11 outbursts considering the uncertainties in the instrumental calibration.
From RXTE archival data, a total of 129 observations span over 389 days in 2002-03, 115 observations over 343 days in 2004-05, 139 observations ranging over 204 days in 2006-07 and 288 observations during 412 days in 2010-11 are considered in our analysis (see Table \ref{outbursts} for details).

HEXTE data are reduced by following standard procedures given in RXTE Cookbook. We have used cluster-A archival data (FS52*.gz) for 2002/03 and 2004/05 outbursts. As cluster-A data did not provide any background measurements, we have used cluster-B archival mode data (FS58*.gz) for 2006/07 outburst. For 2010/11 outburst, we have used cluster-A event mode data (FS50*.gz) for some observations and rest of the observations are excluded due to the problem in rocking motion of HEXTE. Using standard procedures we have extracted background subtracted and dead time corrected source and background spectra as well as the response informations. HEXTE spectrum is generated in the energy range 15.0-200.0 keV without any grouping/binning.

\subsection{Swift Observations}
Swift X-ray telescope (XRT) is a X-ray focusing telescope in the energy range 0.2-10.0 keV.  
GX 339-4 is a bright source and XRT observed the source in windowed timing mode which produces an one dimensional image. We run \enquote{xrtpipeline} to generate level 2 event file and we use \enquote{xselect} to generate the source and background spectrum. We have selected a circle of radius 30 pixel around the image centre for the source region and background from an annular region between 70 and 130 pixel, keeping the same diameter for source and background. For data with pile up we extract source region from an annular region with outer circle of radius 30 pixel and inner radius varies with the degree of pile up. Then we edit the BACKSCAL keyword in the source and background spectrum. We use the response files \enquote{swxwt0to2s0$\_$20070101v012.rmf} for 2006/07 and \enquote{swxwt0to2s6$\_$20090101v015.rmf} for 2010/11 outbursts and generate ancillary response function file using the task \enquote{xrtmkarf}.  The spectrum is generated in the range 0.3-10.0 keV with a grouping of 20 photons per bin without adding systematic. We have analysed swift data in the time period 23-04-2007 to 25-05-2007 for 2006/07 and 21-01-2010 to 19-06-2010 for 2010/11 outbursts (Table \ref{xmm_xrt_pca_obs}) which are simultaneous with the RXTE data.
\begin{table}
\caption{Outbursts and durations}
\begin{tabular}{ c c c c c }
\hline 
No & Start date (MJD) & Stop date (MJD) & Duration (day) \\ 
\hline 
1 & 02-04-2002 (52366) & 27-04-2003 (52756) & 389 \\ 
2 & 13-05-2004 (53138) & 20-04-2005 (53480) & 343 \\ 
3 & 14-11-2006 (54053) & 05-06-2007 (54256) & 204 \\ 
4 & 12-01-2010 (55208) & 28-02-2011 (55620) & 412 \\ 
\hline 
\end{tabular} 
\label{outbursts}
\end{table}
\subsection{XMM-Newton Observations}
The European Photon Imaging Camera (0.2-15.0 keV) in XMM-Newton contains two MOS CCD and one PN CCD. There are four XMM-Newton observations took place simultaneous with the RXTE observations. As MOS data are highly piled up we could not get any useful results from two observations on 24-08-2002 and 19-09-2002. We have extracted the spectra by analysing the observations on 08-03-2003 and 20-03-2003 (Table \ref{xmm_xrt_pca_obs}) from PN timing mode data. The data are reduced using \enquote{SAS version 5.22.2} following the guidelines from cookbook. The events are extracted from a region of ten pixels either side of brightest part of the source. As the source was so bright, there was no source free region to extract a background. To reject bad pixels and events close to chip edges, the events are filtered by specifying \enquote{FLAG=0}, and to select single pixel and double pixel event it requires \enquote{PATTERN$\leq$4}. The extent of pile up on these two observations are found out by using the SAS task \enquote{EPATPLOT}. The effect of pile up is reduced by removing the four pixels for observation on 08-03-2003 and two pixel for observation on 20-03-2003 from the central part. Then ancillary response file is generated using \enquote{arfgen} and the response file is generated by the task \enquote{rmfgen}. The spectrum is generated in the energy range 0.5-8.0 keV by applying a grouping of 20 counts per bin without adding any systematic.

\section{Results and Discussion}
\label{results}
During the life span of RXTE (1995 to 2012), it has observed GX 339-4 to undergo outburst quite frequently. Here we have done a comprehensive study of spectral behaviour of major outbursts of GX 339-4 during this period. Since RXTE did not have the full coverage of 1998/99 outburst, we did not consider this outburst in our study. The four outbursts considered in our study are 2002/03, 2004/05, 2006/07 and 2010/11. The observation details are given in Table \ref{outbursts} and the outburst duration is measured from the starting date given in MJD.
\begin{figure}
\centering
\includegraphics[width=0.5\textwidth]{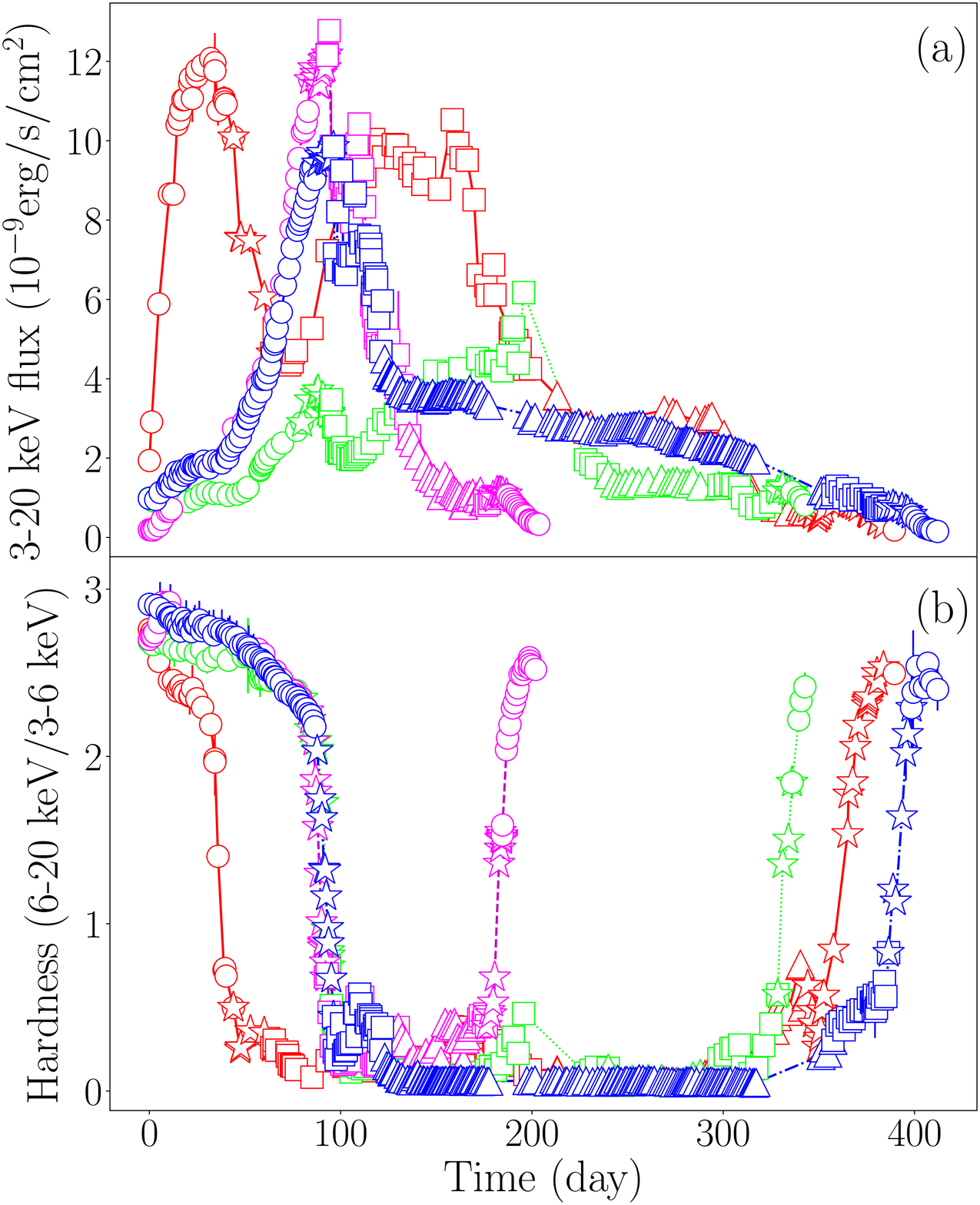} 
\caption{(a) 3.0-20.0 keV PCA light curves and (b) hardness ratio (6.0-20.0 keV flux / 3.0-6.0 keV flux) as a function of time of all outbursts of GX 339-4 during 2002-2011 period are plotted. The lines with different colours (red-solid: 2002/03; green-dot: 2004/05; magenta-dash: 2006/07 and blue-dash dot: 2010/2011) represent individual outbursts. The different symbols: circles mark LHS; star marks HIMS; squares mark SIMS and triangles mark HSS.}
\label{fig:Figure_1}
\end{figure}
\begin{figure}
\centering
\includegraphics[width=0.45\textwidth]{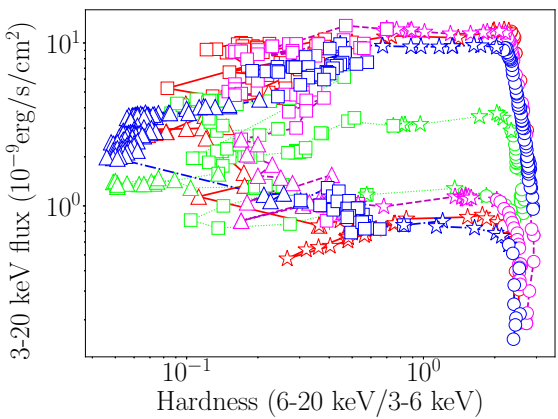} 
\caption{PCA HID of all outbursts of GX 339-4 during 2002-2011 period. The lines of different colours/styles and symbols follow the same meaning as in Figure \ref{fig:Figure_1}.}
\label{fig:Figure_2}
\end{figure}
\begin{figure*}
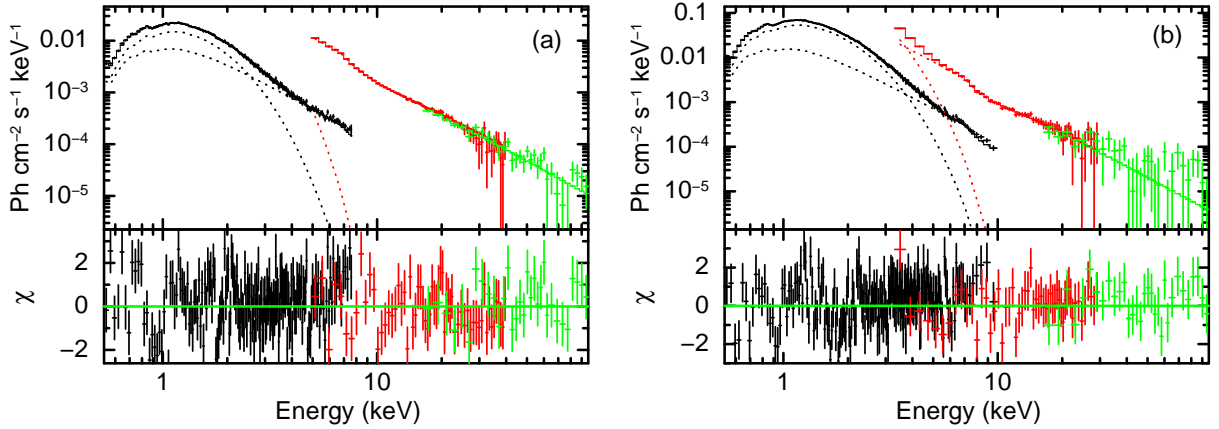

\includegraphics[width=0.32\textwidth,angle =-90]{./figure_3a.eps} 
\includegraphics[width=0.32\textwidth,angle =-90]{./figure_3b.eps}
\caption{Broadband spectral modeling of PN/XMM-Newton (black), PCA (red) and HEXTE (green) observations on (a) 08-03-2003 and (b) 20-03-2003 by phabs*smedge(diskbb+powerlaw)*constant model with $\chi^{2}/dof=1071.56/1000$ and $\chi^{2}/dof=1114.67/1066$ respectively. The PN/XMM-Newton spectrum is binned by a factor of 5 for better clarity.}
\label{fig:Figure_3}
\end{figure*}
\begin{figure*}
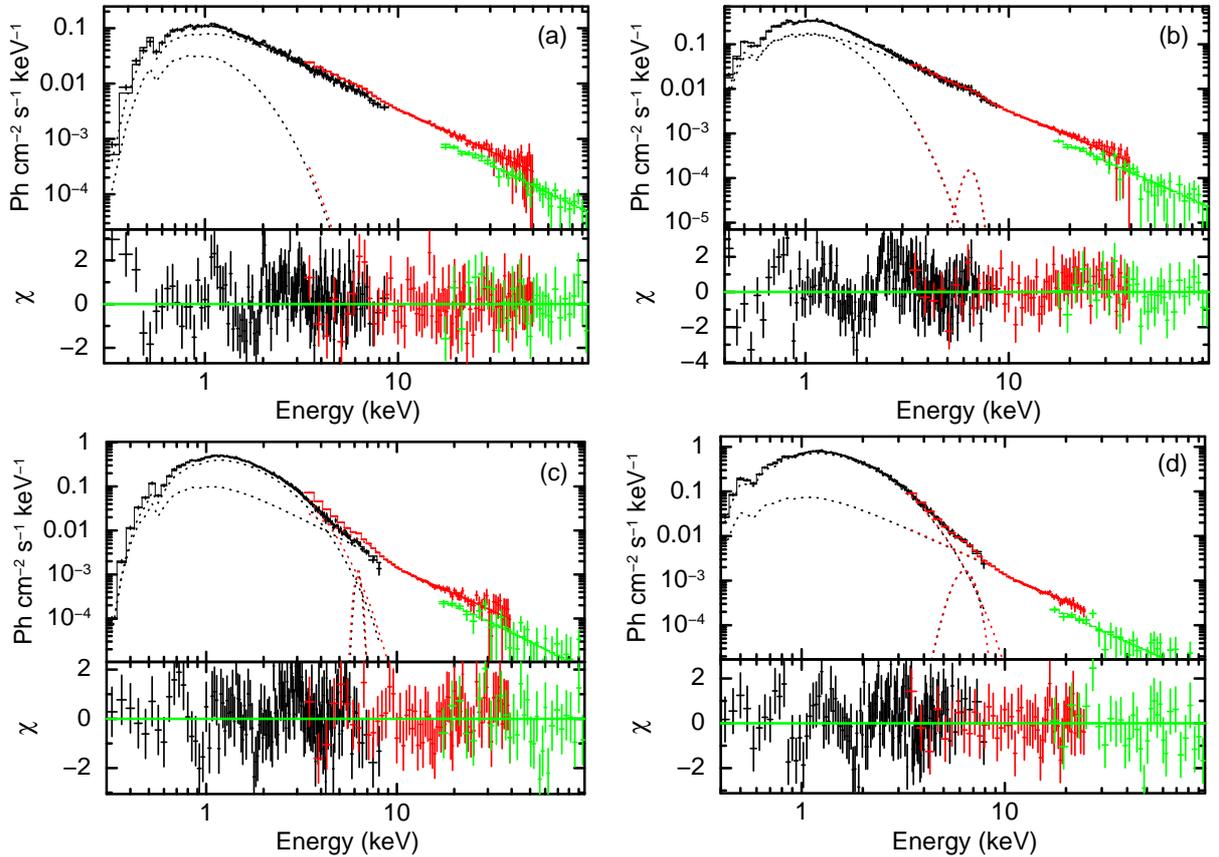

\includegraphics[width=0.32\textwidth, angle =-90]{./figure_4a.eps} 
\includegraphics[width=0.32\textwidth, angle =-90]{./figure_4b.eps}
\includegraphics[width=0.32\textwidth, angle =-90]{./figure_4c.eps} 
\includegraphics[width=0.32\textwidth, angle =-90]{./figure_4d.eps}
\caption{Broadband spectral fitting of XRT/Swift (black), PCA (red) and HEXTE (green) data in (a) LHS (MJD 54240.884), (b) HIMS (MJD 54237.331), (c) SIMS (MJD 54231.913) and (d) HSS (MJD 54213.031) during 2006/07 outburst. The spectra are fitted with phabs*smedge(diskbb+ga+po)*constant model whereas Gaussian component is not required for (a). The fitting statistics $\chi^{2}/dof$ are $549.80/604$, $536.81/583$, $523.86/534$ and $729.44/679$ respectively. The XRT/Swift spectrum is binned by a factor of 4 for better clarity.}
\label{fig:Figure_4}
\end{figure*}
\begin{figure*}
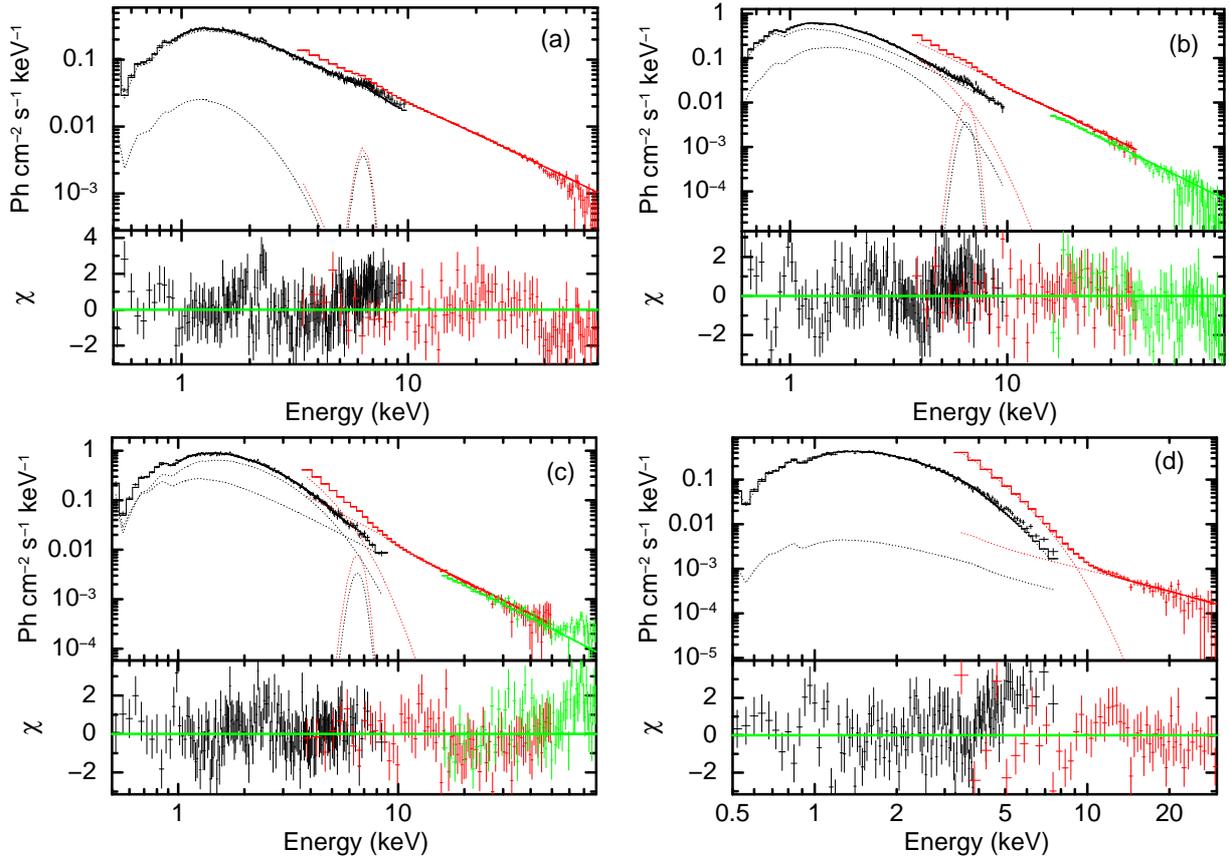

\includegraphics[width=0.32\textwidth, angle =-90]{./figure_5a.eps} 
\includegraphics[width=0.32\textwidth, angle =-90]{./figure_5b.eps}
\includegraphics[width=0.32\textwidth, angle =-90]{./figure_5c.eps} 
\includegraphics[width=0.32\textwidth, angle =-90]{./figure_5d.eps}
\caption{Broadband spectral modeling of XRT/Swift (black), PCA (red) and HEXTE (green) data in (a) LHS (MJD 55281.744), (b) HIMS (MJD 55302.411), (c) SIMS (MJD 55316.986) and (d) HSS (MJD 55334.061) during 2010/11 outburst. The model phabs*edge*smedge(diskbb+gaussian+po)*constant are used to fit the data whereas gaussian is not required for (d). The corresponding statistics for spectral modelling are $\chi^{2}/dof=854.68/819$, $857.36/803$, $727.25/690$ and $643.92/530$ respectively. The XRT/Swift Spectrum is binned by a factor of 4 for better clarity.}
\label{fig:Figure_5}
\end{figure*}

The X-ray fluxes in various energy bands are calculated by fitting PCA spectra using {\it XSPEC version 12.9.1}. We have used a combination of phenomenological models consisting of an absorption component ({\it phabs}) with a {\it powerlaw} and/or a multi-colour disc blackbody ({\it diskbb}), whenever required. Occasionally an emission line represented by a {\it Gaussian} in energy range (6-7) keV, a smeared absorption edge ({\it smedge}) \citep{1994PASJ...46..375E} and high energy cutoff ({\it highecut}) are required to fit the data. We have used photoelectric absorption cross sections from \cite{1996ApJ...465..487V} and elemental abundance \cite{2000ApJ...542..914W} for the absorption model. The chi-square statistic is used for model fitting. In the spectral modelling of all the data set, the best fit model have reduced $\chi^{2}$ between 0.8 to 1.3 and the uncertainities are calculated within 90\% confidence limit using {\it migrad} method. Including all the spectral fitting, the {\it smedge} width varies between 1 to 30keV and the {\it smedge} maximum absorption factor varies between 0.1 to 7 whereas the gaussian line width varies between 0.6 to 1.0.
From spectral modeling, we have calculated the flux in 3.0-20.0 keV energy band to generate the light curve. Also, we have separately estimated the flux in 3.0-6.0 keV and 6.0-20.0 keV energy band and the hardness is defined as the ratio of fluxes in 6.0-20.0 keV and 3.0-6.0 keV. The lightcurves of different outbursts and the corresponding hardness evolution are shown in the top (a) and bottom (b) panels of Figure \ref{fig:Figure_1} respectively.
Here, time of evolution is expressed in day and day zero of individual outbursts in MJD are given in Table \ref{outbursts}. The general evolution of outbursts can understand from HID which have been plotted in Figure \ref{fig:Figure_2}. Here and rest of the paper, we have used lines with different colours (red-solid: 2002/03; green-dot: 2004/05; magenta-dash: 2006/07 and blue-dash dot: 2010/2011) to represent individual outbursts. We follow \cite{2005A&A...440..207B,2011MNRAS.418.2292M,2009MNRAS.400.1603M,2011MNRAS.418.1746S,2012A&A...542A..56N} for spectral classifications and different symbols (circles mark LHS; stars mark HIMS; squares mark SIMS and triangles mark HSS) are used to represent the spectral states.

Generally, in the rising phase of the outburst the flux increases gradually with very small variation in hardness in the low hard state and suffers a sudden drop in hardness in HIMS. The flux reaches maximum value in SIMS where the spectrum becomes softer. The source continues to evolve with low hardness and a decreasing trend of  flux in HSS until it reaches HIMS in decline phase. The hardness evolution shows a very similar trend for all these outbursts even though the duration of outbursts and flux levels are quite different. The hysteresis nature of the HID is ensured by the hard-soft transition at high flux level in the rising phase whereas a soft-hard transition at low flux level in the decline phase. All HIDs (Figure \ref{fig:Figure_2}) have very similar characteristics except 2004/05 outburst being low luminous. 

\subsection{Simultaneous broadband spectral modeling}
\label{simu-pheno}
We have generally modeled the PCA data for spectral study. Also we have searched for nearly simultaneous observations by XMM-Newton and Swift along with RXTE during these outbursts. XMM-Newton and RXTE observations of 2002/03 outburst, Swift and RXTE observations of 2006/07 and 2010/11 outbursts are not strictly simultaneous but still we can generate broadband spectra as the source did not show a large change in count rate as well as hardness during the period of observations. 
The details of these observations are given in Table \ref{xmm_xrt_pca_obs}. The simultaneous broadband spectra (0.3 - 100.0 keV) are modeled using a {\it powerlaw} and a multi-colour disc blackbody ({\it diskbb}). Also, additional components like emission line ({\it gauss}) in energy range (6-7) keV, smeared absorption edge ({\it smedge}) \citep{1994PASJ...46..375E}
 are occasionally used to fit the data. The broadband spectra are required to constrain the blackbody component (particularly in LHS) and a better estimation of powerlaw component as well. The interstellar absorption $n_{H}$ is frozen at $5\times10^{21} cm^{-2}$ \citep{1986A&A...164...67I,1997ApJ...479..926M,2000MNRAS.312L..49K}. Otherwise $n_{H}$ varies between $(4-7)\times10^{21} cm^{-2}$ for broadband spectra using XRT, PCA and HEXTE. 

We could find only two simultaneous observations of PN/XMM-Newton along with RXTE for 2002/03 outburst and the broadband spectra are presented in Figure \ref{fig:Figure_3}. Both observations are in the HIMS state during the decline phase of the outburst. Here black, red and green colours represent the XMM-Newton, PCA and HEXTE data respectively along with unfolded models of {\it diskbb} and {\it powerlaw}. No simultaneous observation exists for 2004/05 outburst. Simultaneous observations of XRT/Swift and RXTE exist during 2006/07 decline phase and rising phase of 2010/11 outbursts.
We have fitted all of them but for representation, we show only one spectrum from each spectral state [(a) LHS; (b) HIMS; (c) SIMS; (d) HSS] for both the outbursts in Figure \ref{fig:Figure_4} and Figure  \ref{fig:Figure_5} respectively.
 In both figures, black, red and green colour represent XRT, PCA and HEXTE data respectively along with unfolded models.

\subsection{Evolution of spectral parameters}
\label{evo-para}
We study the time evolution of spectral fitting parameters using PCA data as well as simultaneous observations (Table \ref{xmm_xrt_pca_obs}). The parameters from PCA data only are shown in green-solid symbols whereas the same from PN-PCA-HEXTE broadband spectra of 2002/03 outburst are shown in blue-dot symbols. Parameters from XRT-PCA-HEXTE spectra for 2006/07 and 2010/11 outbursts are presented by red-dash symbols. We have used the same symbols for spectral classes as in Figure \ref{fig:Figure_1}.

\begin{figure}
\centering
\includegraphics[width=0.47\textwidth]{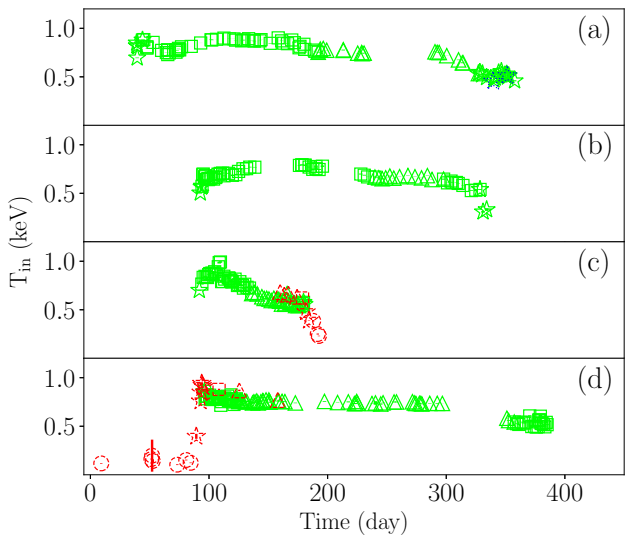}  
\caption{Evolution of inner accretion disc temperature ($T_{in}$) of the source for (a) 2002/03 (b) 2004/05 (c) 2006/07 and (d) 2010/11 outbursts spectral fitting. The green-solid symbols represent only PCA data, blue-dot symbols are for simultaneous XMM-Newton-PCA-HEXTE data and red-dash symbols mark simultaneous XRT-PCA-HEXTE data. The different symbols: circles mark LHS; star marks HIMS; squares mark SIMS and triangles mark HSS.The uncertainities (marked inside the symbols) are within 90\% confidence range.}
\label{fig:Figure_6}
\end{figure}

\begin{figure}
\centering
\includegraphics[width=0.47\textwidth]{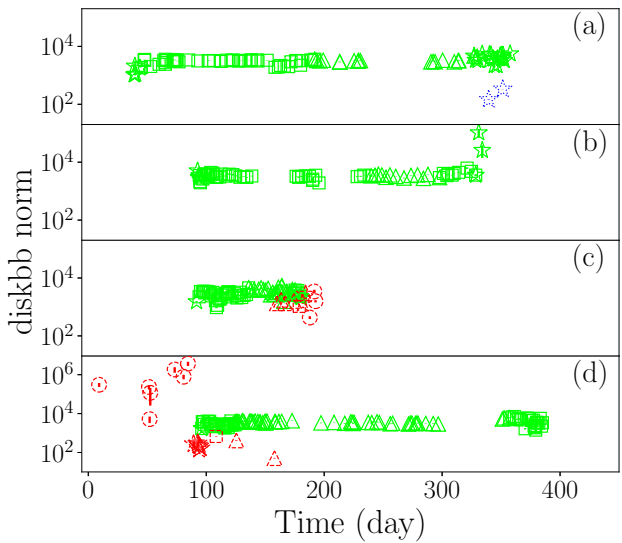} 
\caption{Evolution of {\it diskbb} normalizations for (a) 2002/03  (b) 2004/05 (c) 2006/07 and (d) 2010/11 outbursts. The different colours and symbols have same meaning as in Figure \ref{fig:Figure_6}. The uncertainities (marked inside symbols) are within 90\% confidence range.}
\label{fig:Figure_7}
\end{figure}

In Figure \ref{fig:Figure_6}, we have plotted the evolution of inner disc temperature $T_{in}$ of {\it diskbb} model for (a) 2002/03, (b) 2004/05, (c) 2006/07 and (d) 2010/11 outbursts respectively. In the LHS, fitting only PCA data require unusually high disc temperature $T_{in} \ge 2 keV$  (see also \citet{2014MNRAS.442.1767P}) and hence only PCA data cannot constrain $T_{in}$ in LHS. In fact, we fit the LHS PCA data with {\it powerlaw}, {\it smedge}. Simultaneous observations with XRT-PCA (red-dash points in Figure \ref{fig:Figure_6}c and Figure \ref{fig:Figure_6}d) could able to estimate $T_{in}$ in LHS. Whereas the simultaneous fitting with PN-PCA data in 2002/03 (blue-dot points in Figure \ref{fig:Figure_6}a) are in HIMS and show similar $T_{in}$ as in PCA. 
This figure shows that in the starting and ending part of the outburst, there may be a small contribution of photons from the accretion disc and hence one expects a low $T_{in}$. As the source approaches to rising-intermediate state, the disc moves inward and $T_{in}$ increase. It remains almost constant in SIMS to HSS and towards the end of declining phase disc starts disappearing with a decreasing $T_{in}$. Also, 2006/07 outburst shows a faster evolution of $T_{in}$ in compared with other outbursts. The evolution of diskbb normalizations are shown in Figure \ref{fig:Figure_7}. The broadband fitting (red-dash points in Figure \ref{fig:Figure_7}) show that disc norm start with a high value in LHS as the disc is far apart and decreases towards intermediate states.
  
\begin{figure}
\centering
\includegraphics[width=0.47\textwidth]{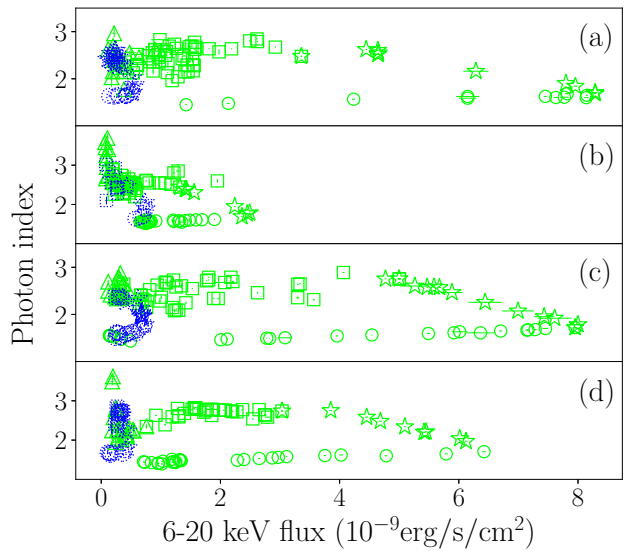} 
\caption{Photon index vs 6.0-20.0 keV flux for (a) 2002/03  (b) 2004/05 (c) 2006/07 and (d) 2010/11 outbursts. The different symbols have same meaning as in Figure \ref{fig:Figure_6}. The green-solid symbols represent the rising phase and blue-dot symbols denote the decline phase of the outbursts. The uncertainities are within 90\% confidence range.}
\label{fig:Figure_8}
\end{figure}

\begin{figure} 
\centering
\includegraphics[width=0.47\textwidth]{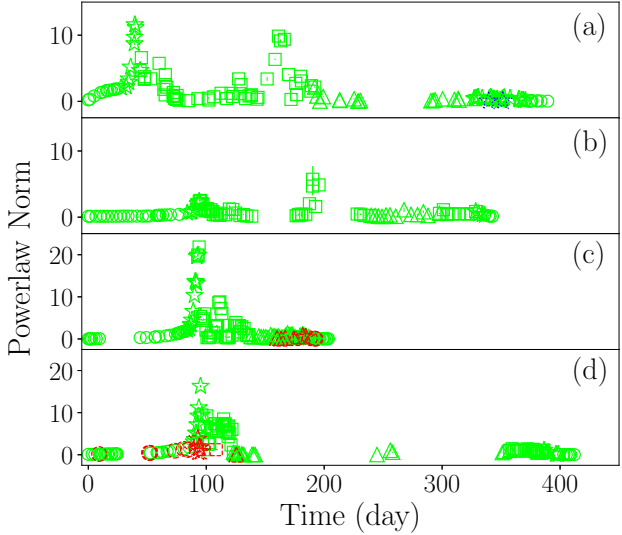} 
\caption{Evolution of powerlaw normalizations for (a) 2002/03  (b) 2004/05 (c) 2006/07 and (d) 2010/11 outbursts. The different colours and symbols have same meaning as in Figure \ref{fig:Figure_6}. The uncertainities are within 90\% confidence range.}
\label{fig:Figure_9}
\end{figure}

The variation of photon index ($\Gamma$) with the hard flux (6.0-20.0 keV) is presented in Figure \ref{fig:Figure_8}. Here, green-solid symbol represents the rising phase whereas blue-dotted symbol denotes the decline phase of the outbursts. All the outbursts show a hysteresis (anti-clock wise) with almost constant photon index ( $\sim 1.5$) in the rising LHS. The value of photon index increases ($2<\Gamma<3$) with a decrease of hard flux as the source moves towards softer states. In the decline part of the outburst photon index decreases and the 6.0-20.0 keV flux does not show much variation. \cite{2004MNRAS.351..791Z}, showed a similar behaviour between photon index and flux in different energy range for the same source. Whereas \cite{2015ApJ...813...84G}, found  nearly constant photon index in hard state of different outburst of GX 339-4 with very different luminosities. 
 The Figure \ref{fig:Figure_9} represents evolution of powerlaw normalization and it shows that the powerlaw normalisation sharply increases towards HIMS of the rising phase where powerlaw dominates the spectra and after that almost remain at constant low values in the softer states. In all outbursts, we see a hard flare in SIMS (parameter $t_{6-20}$ in Table \ref{lcpeak}) indicated by a sudden rise of powerlaw normalization. We find that a powerlaw component is always required to model data over the entire outburst whereas in 2010/11 outburst (Figure \ref{fig:Figure_9}d) the spectral modeling in HSS can be done using {\it diskbb} only. 
 
\begin{figure}  
\centering
\includegraphics[width=0.47\textwidth]{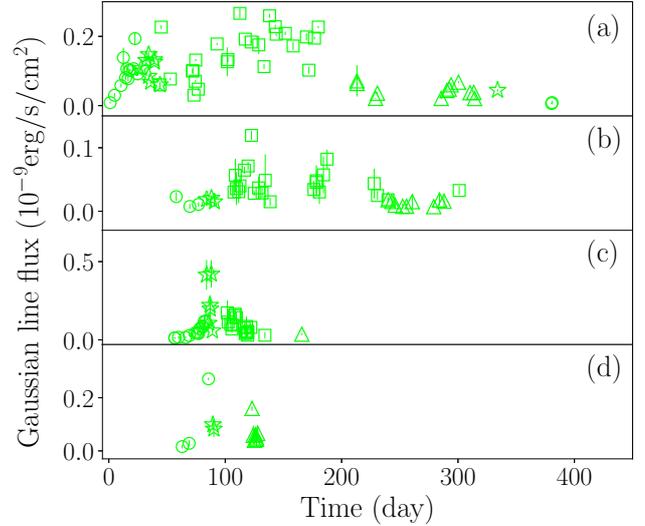} 
\caption{Evolution of gaussian line flux during (a) 2002/03  (b) 2004/05 (c) 2006/07 and (d) 2010/11 outbursts. We follow the same representations as in Figure \ref{fig:Figure_6}. The uncertainities are within 90\% confidence range.}
\label{fig:Figure_10}
\end{figure}

\begin{figure}  
\centering
\includegraphics[width=0.47\textwidth]{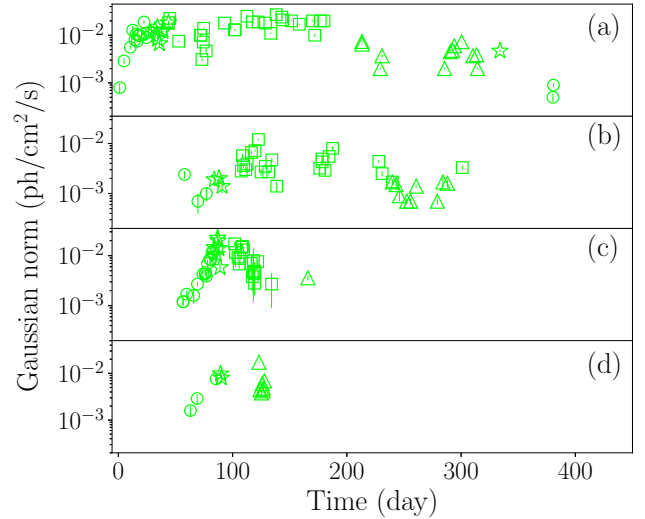} 
\caption{Evolution of normalisation of gaussian line during (a) 2002/03  (b) 2004/05 (c) 2006/07 and (d) 2010/11 outbursts. We have used the same symbols and colour references (Figure \ref{fig:Figure_6}). The uncertainities are within 90\% confidence range.}
\label{fig:Figure_11}
\end{figure}

Past several studies suggested that Fe line flux can be used as a tracer of Compton reflection signature. The process of bound-free absorption in the reflecting medium is followed by emission of a fluorescent Fe K$\alpha$ line. The equivalent width of the Fe K$\alpha$ line is a measure of the extend of reflection. \cite{2001MNRAS.328..501P} discussed how Comptonization process affects the measurement of the equivalent width of Fe line and showed that the unscattered equivalent width of Fe line is anti-correlated with the corona temperature. In the same sense, \cite{2003MNRAS.342..355Z} found Compton reflection strength is correlated with photon index for many outbursting sources and is strongest in the soft state whereas \cite{2014SSRv..183..121G} showed a  correlation between photon index and the relative strength of the Comptonized radiation. \cite{2016ApJ...829L..22S} presented strong anti-correlation between Fe line equivalent width with powerlaw normalization. 
We have used Fe line represented by {\it gauss} to fit the data whenever required. In Figure \ref{fig:Figure_10}, we have plotted the evolution of Gaussian line flux and it shows a very interesting trend. We observe that the Compton reflection signature is week in LHS and has an increasing trend of reflection strength towards softer states. Finally, becomes strongest in the SIMS for all outbursts except 2010 outburst. Also, if the existence of Fe $K\alpha$ line is a signature of reflection, we observe a decreasing trend of reflection signature required to model the successive outbursts after 2002. To strengthen this findings,  
in Figure \ref{fig:Figure_11}, we have plotted the evolution Gaussian line normalization over time with its uncertainty. We hardly could see the presence of Gaussian line in LHS except 2002/03 outburst. In LHS, we could able to fit the data reasonably well with reduced $\chi^{2}$ in the range 0.8-1.1.  Also, we see from Figure \ref{fig:Figure_11} that in 2002/03 outburst the minimum value of line normalization is as low as $10^{-4}$ and it increases in successive outbursts. For 2010/11 outburst the minimum normalization is above $10^{-3}$. Possibly, this increasing trend of minimum normalization of Gaussian line is an indication of the degradation of PCA energy response over time. Though this decreasing trend of the requirement of Gaussian line may be due to the physical differences between the outbursts as well. In fact the reason that we observe outbursts of different durations, peak luminosities etc are the reflection of physical differences between outbursts. But in that case one should not expect any particular trend between outbursts.
 


\subsection{Spectral modelling using two-component accretion flow model}
\label{simu-phy}
\begin{figure}
\includegraphics[width=0.35\textwidth,angle =-90]{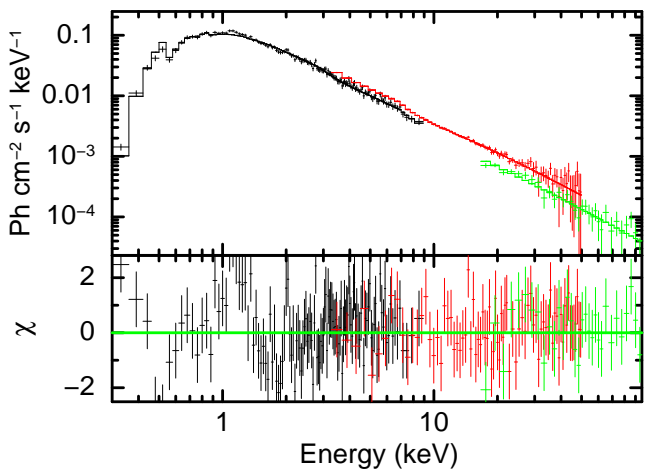} 
\caption{A representation of broadband spectral fitting of XRT/Swift (black), PCA (red) and HEXTE (green) data in LHS (MJD 54240.884) using two-component model with additional {\it smedge} component. The XRT/Swift spectrum is binned by a factor of 4 for better clarity. The statistics of the fitting is $\chi^{2}/dof=574.18/601$. }
\label{fig:Figure_12}
\end{figure}

In \S\ref{evo-para}, we have studied the evolution of spectral parameters by fitting the data with phenomenological models where the spectral parameters are not directly connected to the hydrodynamics and radiation processes of the system. Hence, it is not possible to  understand the behaviour of dynamical parameters from phenomenological spectral parameters. We intend to study the evolution of the dynamical parameters by fitting the spectral data with a physical model. A two-component advective accretion flow model \citep{1995ApJ...455..623C,2006ApJ...642L..49C} provides a natural explanation for the two distinct components of the radiation spectrum, namely the soft part is associated with a standard Keplerian accretion disc and the hard part comes from a hot sub-Keplerian plasma cloud. Also, the need for two distinct components are well supported by the observed timing properties \citep{2001ApJ...554L..41S,2002ApJ...569..362S,2007ApJ...669.1138S} of many galactic black hole candidates. In this model a standard Keplerian accretion disc resides on the equatorial plane and sub-Keplerian halo occupies above and below the standard disc.
A Keplerian flow cannot enter to the black hole (BH) since the inner boundary condition demands a supersonic flow close to horizon. Hence, the flow close to the BH is essentially sub-Keplerian in nature. Since the sub-Keplerian halo component also possesses significant angular momentum, it creates a centrifugal barrier close to BH which slows down the inflowing matter. In fact, the centrifugal barrier may lead to a shock transition as well for the right choice of input parameters, like energy, angular momentum of the flow. This enhances the density and temperature of the central sub-Keplerian flow i.e., post-shock region (POSR) which serves the purpose of the so-called `Corona' \citep{1974ApJ...187L...1L,1975ApJ...199L.153E,1976ApJ...204..187S,1993ApJ...413..507H}. Hence, POSR is an integral part of sub-Keplerian halo. Only one expects a change in geometry of the sub-Keplerian flow at shock: POSR is more of tapered or spherical shape due to shock compression whereas pre-shock flow has a geometrically thick disc structure. Rather the model assume Keplerian disc is truncated at the shock \citep{2015MNRAS.448.3221G,2016ApJ...826...23T}. The Keplerian disc produces multi-colour black body photons and a fraction (typically around 10-15\% depending on the geometry of POSR) of soft photons is intercepted by POSR which inverse-Comptonized these soft photons to produce the powerlaw high energy photons. For a given mass and accretion rate ($\dot m_d$), we calculate the radiation spectra from individual annuli of Keplerian disc following the standard disc prescription \citep{1973A&A....24..337S} and add all the contributions to find multi-colour blackbody spectrum from the Keplerian disc. We calculate the radial distribution of flow variables (density and temperature distributions) in the sub-Keplerian flow by solving the hydrodynamic equations including inverse-Compton cooling in the POSR. We self-consistently calculate the fraction of soft photons from Keplerian disc intercepts the POSR. Using these flow variables and intercepted soft photon fraction, we calculate the Comptonized radiation spectrum from POSR following \cite{1995ApJ...455..623C}. Finally, the overall spectrum is a combination of the multi-colour blackbody spectrum from Keplerian disc and the inverse-Comptonized spectrum from POSR. Thus the spectral fitting Figure \ref{fig:Figure_12} using two-component model shows only one component i.e., the overall spectrum and no separate blackbody component or inverse-Comptonized component.

 In this model, there are five free parameters, namely, the mass of the black hole ($M$), two accretion rates (Keplerian accretion rate $\dot m_d$ and sub-Keplerian accretion rate $\dot m_h$), the size of the POSR ($x_s$) and the overall normalization. Here, mass is expressed in unit of $\textup{M}_\odot$, accretion rates are expressed in  Eddington rate ($\dot M_{Edd}=1.4 \times 10^{18} (M/\textup{M}_\odot)$ gm/s with 10\% efficiency of accretion) and distance is expressed in unit Schwarzschild radius $r_g = 2GM/c^2$, where $G$ is universal gravitational constant and $c$ is speed of light in vacuum. 
 This model has been implemented in {\it xspec} as a local table model \citep{2014MNRAS.440L.121D,2015ApJ...807..108I}. We fit all the PCA data  during 2002-2011 following Iyer et al., 2015. Here we assume the mass of the source as $9 \textup{M}_\odot$ \citep{2016ApJ...821L...6P} and we have fixed $n_{H}=5\times 10^{21} cm^{-2}$ \citep{1986A&A...164...67I,1997ApJ...479..926M,2000MNRAS.312L..49K} for the entire spectral modeling. Also we see occasional requirement of additional components like {\it gauss} and {\it smedge} along with two-component model as those physics are yet to be included into the model. We have used photoelectric absorption cross sections \citep{1996ApJ...465..487V} and elemental abundance \citep{2000ApJ...542..914W} for absorption model. We have used chi-square statistic to represent the goodness of fit. We have modelled the PCA data of all the four outbursts and the best fit reduced $\chi^{2}$ varies between 0.8 to 1.3. The uncertainities are calculated within 90\% confidence range using method {\it migrad}. We find that the smearing width varies between 2-25 keV, the {\it smedge} maximum absorption factor varies between 0.1 to 5 whereas the gaussian line width varies between 0.4 to 1.0. As a representation, in Figure \ref{fig:Figure_12} we have fitted the broadband spectra of XRT/Swift (black), PCA (red) and HEXTE (green) data in LHS (MJD 54240.884) using two-component model with additional {\it smedge} component. The two-component model parameters are $\dot m_h=0.016\pm0.002$, $\dot m_d=0.009\pm0.002$, $x_s=68\pm5.6$ with $\chi^2/dof=574.28/601$.

\begin{figure*}
\centering
\includegraphics[width=1.0\textwidth]{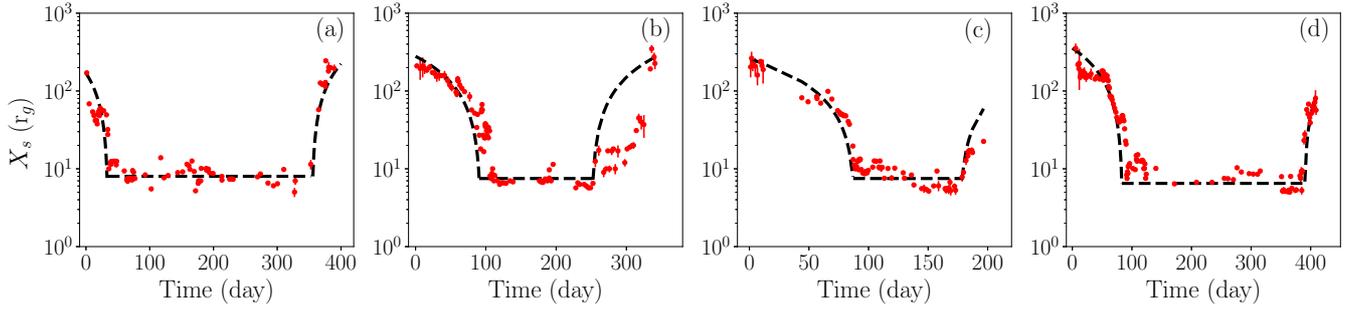}

\caption{Evolution of shock location (red points) for (a) 2002/03, (b) 2004/05, (c) 2006/07 and (d) 2010/11 outbursts. The black dashed line represents the fitting of the general behaviour of $x_s$ using Equation \eqref{eq:xs}.}
\label{fig:Figure_13}
\end{figure*}
\begin{figure*}
\centering
\includegraphics[width=1.0\textwidth]{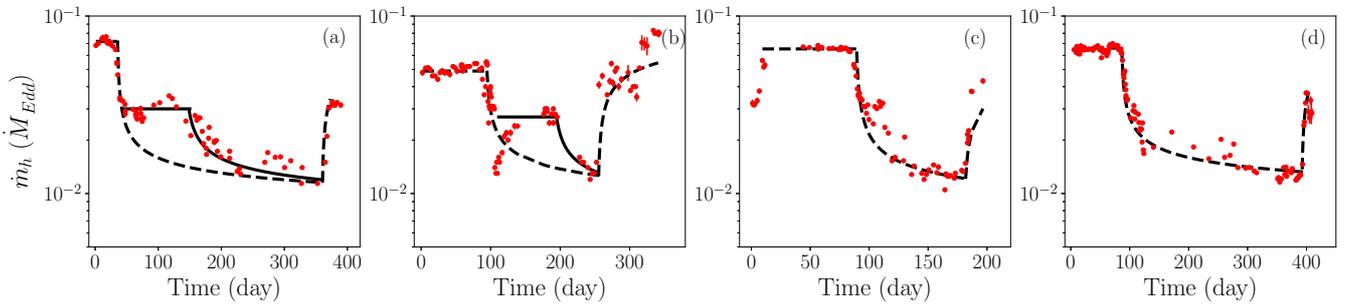}
\caption{Evolution of sub-Keplerian halo rates (red points) for (a) 2002/03, (b) 2004/05, (c) 2006/07 and (d) 2010/11 outbursts. The black dashed line represents the fitting of the general behaviour of $\dot m_h$ using Equation \eqref{eq:mh}. The solid lines in (a) and (b) appear to be a second triggering of outburst (see \ref{unified} for details).}
\label{fig:Figure_14}
\end{figure*}
\begin{figure*}
\centering
\includegraphics[width=1.0\textwidth]{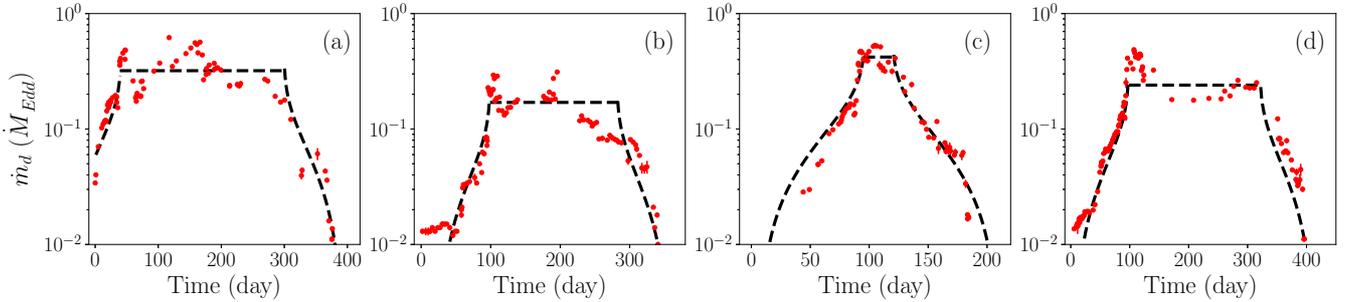}   
\caption{Evolution of Keplerian disc rates (red points) for (a) 2002/03, (b) 2004/05, (c) 2006/07 and (d) 2010/11 outbursts. The black dashed line represents the fitting of the general behaviour of $\dot m_d$ using Equation \eqref{eq:md}.}
\label{fig:Figure_15}
\end{figure*}

The time evolution of the model parameters using all four outbursts data are presented in Figure \ref{fig:Figure_13},  Figure \ref{fig:Figure_14} and Figure \ref{fig:Figure_15} respectively. Here red points are parameters values from spectral modeling of individual data set using two-component flow. The day zero of each outburst corresponds to MJD given in Table \ref{outbursts} and panels (a, b, c and d) represent the 2002/03, 2004/05, 2006/07 and 2010/11 outburst respectively. The general behaviour of parameters ($x_s, \dot m_h$ and $\dot m_d$) are very similar (except $\dot m_h$ in Figure \ref{fig:Figure_14}a,b and is discussed in \S\ref{accpara}) in all four outbursts and we can understand the spectral evolutions by the relative values of these parameters. 
At the beginning of the outburst $\dot m_h$ is high (Figure \ref{fig:Figure_14}) and the shock location (Figure \ref{fig:Figure_13}) is far away from the central object. Hence, POSR contains a large number of high energy electrons but does not have sufficient soft photons to cool the POSR as the Keplerian disc rate (Figure \ref{fig:Figure_15}) is very low. Hence, the source is in LHS. As the source proceeds towards the peak of the rising phase, $x_s$ moves inward, $\dot m_h$ starts decreasing sharply (indicates no more fresh matter supply at the outer edge), $\dot m_d$ increases. This enhances the supply of soft photons and POSR becomes smaller and the source makes a transition to intermediate states. This increases the overall luminosity. The shock location remains closer to the central object, almost at constant value, during SIMS to HSS and $\dot m_h$ keep on decreasing whereas $\dot m_d$ remains around a constant high value. In HSS state, $\dot m_d$ starts decreasing and $\dot m_h$ has a very low value which makes hardness to a lower value. At the end shock location starts moving outwards, $\dot m_h$ starts increasing again, $\dot m_d$ continue to decrease to a low value and the source moves back to hard state in the declining phase. This explains the hysteresis behaviour of HID (Figure \ref{fig:Figure_2}) as well.

\subsection{Modeling the evolution of accretion parameters}
\label{accpara}

Just like dwarf novae outburst, the transient low mass X-ray binary (LMXB) also shows recurrence outbursts. The key difference between the two outbursts is that the dwarf novae show repeated periods of optical, UV outbursts last for days separated by quiescence of several weeks whereas transient LMXB undergoes mostly X-ray outbursts last for many days/months and the recurrence is neither so frequent nor periodic as in dwarf nova. It is generally accepted that the outburst in dwarf nova can be understood by the thermal-viscous disc instability model \citep[DIM:][for a review]{1974PASJ...26..429O,2001NewAR..45..449L}. 

The disc instability model (DIM) can be put forward in two-component flow model picture as: during quiescence, the accretion to the central object from the outer boundary happens through sub-Keplerian flow and the Keplerian disc is truncated out to a distance of few hundred $r_g$ (parameter $Xs_0$ in Table \ref{fit_par}) from the central BH. As the outburst triggers due to DIM the Keplerian disc moves inward and more matter is accreted by the Keplerian disc. This increases the soft photons flux as well as hard photons flux due to inverse Comptonization of intercept soft photons by the POSR i.e., the overall flux increases. Also during this phase a constant sub-Keplerian rate supplement the increase of hard photons as the intercepted Keplerian soft photons are not enough to cool the POSR. Around the peak of the outburst, the central X-ray source could able to irradiate the outer part of the disc which enhances the outburst duration by transferring the outer Keplerian disc into hot state. Since the sub-Keplerian flow resides on top/bottom of the Keplerian disc, the enhanced viscosity at the outer part due to irradiation could able to convert the sub-Keplerian matter into Keplerian matter and hence sub-Keplerian rate starts decreasing. This decreases the hard flux and the source moves into a quasi steady accretion phase as long as Keplerian accretion rate is approximately constant and is above the critical accretion rate required to maintain the hot branch. The source luminosity decreases as almost all the Keplerian matter accreted by the BH and the Keplerian disc recedes back towards outer edge to a cool state. The low value of viscosity at the outer edge allows the sub-Keplerian matter to rebuild in the accretion and bring back the source into a quiescence state. We see a replication of this scenario from the evolution of the accretion parameters (Figure \ref{fig:Figure_13}, Figure \ref{fig:Figure_14} and Figure \ref{fig:Figure_15}).


In this study, we are trying to understand the general behaviour the evolution of the accretion parameters using a toy model. We define four time scales: $t_{r}$ represents the rising time of the soft flux (related to $\dot m_d$), $t_h$ is the rising time of hard flux (related to $m_h$), $t_{d}$ is the time scale after which $\dot m_d$ starts decreasing in the decline phase and $t_{s}$ is the time scale that both $\dot m_h$ and $x_s$ begin to increase in the decline phase of the outburst.

Accordingly, the Keplerian accretion rate evolves with time as,
\begin{equation}
\dot{m}_{d}= 
\begin{cases}
\begin{split}
A-\alpha \  log(t_{r}-t),\qquad\qquad t<t_{r} \\
A, \qquad\qquad t_{r}<t<t_{d}\\
A-\alpha \ log(t-t_{d}),  \qquad\qquad t>t_{d}  
\end{split}
\end{cases}     
\label{eq:md}
\end{equation}
where $A$ is the highest average value of the Keplerian rate in unit of Eddington rate, $t$ is time in days and $\alpha$ is a free parameter defines the steepness of the evolution. 

The sub-Keplerian accretion rate can be modeled as,
\begin{equation}
\dot{m}_{h}=
\begin{cases}
\begin{split}
B,  \qquad\qquad\qquad  t<t_{h} \\
\frac{B}{\beta} \times \frac{1}{log(t-t_{h})},  \qquad\qquad  t_{h}<t<t_{s}  \\
\frac{B}{\beta} \times  log(t-t_{s}),  \qquad\qquad  t>t_{s}
\end{split}
\end{cases}  
\label{eq:mh}      
\end{equation}
where $B$ is the maximum value of $\dot m_h$ in Eddington rate, $\beta$ is a free parameter and $t$ is time in days.

The shock location varies as \citep{2010ApJ...710L.147M} 
\begin{equation}
x_{s}=
\begin{cases}
\begin{split}
X_{s_{0}}-v_{0}t, \qquad\qquad t<t_{h} \\
X_{s_{i}}, \qquad\qquad t_{h}<t<t_{s}  \\
X_{s_{i}}+v_{0}(t-t_{s}),  \qquad\qquad  t>t_{s}
\end{split}  
\end{cases}
\label{eq:xs}         
\end{equation}
where, $X_{s_{0}}$ is the outer most value of shock location at the triggering of the outburst, $v_{0}$ is the constant velocity of shock front per day and $X_{s_{i}}$ is inner most position of the shock location. 

We have fitted the data (red points) in Figure \ref{fig:Figure_13} by Equation \eqref{eq:xs}, data presented in Figure \ref{fig:Figure_14} by Equation \eqref{eq:mh} and parameter values in Figure \ref{fig:Figure_15} using Equation \eqref{eq:md} respectively without much attention on the local variations. The general profiles are represented by black dashed lines in Figures \ref{fig:Figure_13}, \ref{fig:Figure_14}, \ref{fig:Figure_15} respectively and the corresponding fitting parameters values are presented in Table \ref{fit_par}. 
We see that $t_r$ and $t_h$ represent the rising time of soft ($t_{3-6}$) and hard flux ($t_{6-20}$) in Table \ref{lcpeak} whereas $t_s$ represents beginning of HIMS in the decline phase. But $t_d$ represents a time scale when the source in SIMS or HSS and does not have very defined identification like other time scales. The overall luminosity depends on the Keplerian accretion rate and 2006/07 outburst is the brightest with highest values of $A$. The maximum average values $\dot m_h$ ($B$) are very similar in all outburst though little lower for 2004/05 outburst whereas the same for $\dot m_d$ ($A$) are very different between outbursts. Hence, the viscous characteristics of the Keplerian disc are very different between outbursts and produce different peak luminosities. In fact, the overall durations of outbursts are very different and it depends on the amount of time the source spends in SIMS and HSS. This in turns depends on the maximum Keplerian accretion rate and the efficiency of viscous dissipation converting it into radiation. Thus in 2006/07 outburst $\dot m_d$ has the maximum value with maximum overall flux among all four outbursts and hence a shorter duration of outburst with short SIMS and HSS. This is possibly due to faster consumption of the Keplerian disc mass in comparison with other outbursts.
The value of $\alpha$ and $\beta$ depend on how accretion rates need to be increased with time to fit the spectra. In the case of GX 339-4, they are found to be the same in both the rising and decay profile of the accretion rates. This is because the rise and decay duration of accretion rates are approximately the same in all the outbursts. But this may not be a general conclusion valid for all outbursts of similar types. Also, a larger values of $\beta$ indicates a sharp rise and decline of $\dot m_h$ than $\dot m_d$. This is expected because $\beta$ depends on the infall time scale (via radiative cooling time scale) whereas $\alpha$ depends on the viscous time scale. In general the shock may propagate in constant speed or accelerated motion due to the inverse Compton cooling of POSR by the soft photons from the Keplerian disc. \cite{2013MNRAS.431.2716M} have calculated the amount of cooling of POSR due to inverse-Comptonization with the increase of Keplerian rate and have shown that shock moves inward with the increase of Keplerian rate. Hence, the shock travels at speed determined by the rate of change of $\dot m_d$ and a typical value is few rg/day. \cite{2018Ap&SS.363...90N} have shown that an accelerated motion of the shock is required to model the QPO evolution of 1999 outburst of XTE J1859+226. But from Table \ref{fit_par} we could see for GX 339-4 the shock front moves inward and outward with constant speed and are very similar between outbursts. Also, the constant value ($v_{0}$) of shock speed may differ during the rising and decay of outbursts.  

The shock location in two component model is the truncation radius of the Keplerian disc. We find from the spectral modeling of all the outburst data that the minimum value of the shock location is $\sim 5 r_{g}$ in the soft state. Although it can be even lower depending on the position of the multiple sonic points of the flow. In general, one expects the truncation radius to move inwards due to the progress of the outburst, being close to the central object in the soft state and move apart in the decline phase. Hence, in the hard state, the truncation radius of the Keplerian disc expected to be far from the BH. Although, \cite{2015ApJ...813...84G} indicated a huge disagreement of the inner edge radius from the reflection modelling of different observations data in the hard state of this source. This could be due to the inherent dependency of inclination angle in the reflection models to determine the truncation radius of the disc. Also, in LHS, the corona contribution is dominant and generally the reflection signature is weak. Hence, a proper continuum modeling of the data is required to constrain the reflection contribution. The continuum contributions from variety of corona models may differ significantly and may result different reflection contributions. Hence, the disc truncation radius calculated using reflection modeling may suffer some uncertainty due to corona models as well.
\begin{table}
\caption{Two component model fitting parameters}
\resizebox{\columnwidth}{!}{%
\begin{tabular}{|c|c|c|c|c|c|c|c|c|c|c|c|}
\hline 
Outbursts &  \multicolumn{4}{|c|}{Time (day)} &$\alpha$& $A$& $B$&$\beta$& $X_{s_{0}}$ & $X_{s_{i}}$ & $v_{0}$ \\
• & $t_{r}$& $t_{h}$ & $t_{d}$ & $t_{s}$ & • &• & •&  •& ($r_{g}$) & ($r_{g}$) &($r_{g}/day$)\\ 
\hline
2002/03 &41.0&33.0& 299.6 & 357.0 &1.63 & 0.32 &  0.072 &2.4& 169 & 8.0 & 4.89 \\ 
2004/05& 99.0&91.0 & 282.0 & 252.5 & 0.91 &0.17&  0.049 &1.75 &279& 7.5 & 3.00 \\ 
2006/07& 96.0&87.0 & 120.0& 178.6&2.15 & 0.42 & 0.065&2.7 &259 & 7.5 & 2.89\\  
2010/11 & 98.0&83.0 & 321.0& 390.1& 1.22  & 0.24 & 0.066 &2.0 &355 & 6.5 & 4.20\\ 
\hline 
\end{tabular}%
}
\
\label{fit_par}
\end{table}
The rising phase of the outburst is controlled by two time scales ($t_{h}$ and $t_{r}$) and one expects $t_{h}<t_{r}$ (see Table \ref{fit_par}) as the hard flux leading the soft flux (Figure \ref{fig:Figure_A1}) whereas $t_{d}$ and $t_{s}$ regulate the decline phase. In general $t_{d}<  t_{s}$ i.e., the Keplerian disc rate starts to decrease before shock location could move outward and the increase of halo rate to initiate the dimming of outburst. This is what we see from Table \ref{fit_par} for  all  outbursts except 2004/05 outburst. This could be the reason that Figure \ref{fig:Figure_14}b  shows deviation from the model trend in the decay phase. In this case the shock start to recede back in the soft state itself before the declination of the Keplerian accretion rate.

The outburst may trigger when the surface density exceed the critical value due to thermal-viscous instability and the associated propagation of heat front, transform the disc into a hot disc with an increase of accretion rate. The standard picture of irradiated DIM \citep{1994A&A...290..133V,1998MNRAS.295L...1S} may be applicable to GX 339-4. We have calculated the outer radius \citep{1986PhT....39j.124F} of the Keplerian disc, $R_{0} \sim 4.4\times 10^{11}$ cm (given $M \sim 9.0 \textup{M}_\odot$, mass ratio $q \sim 0.18$ and binary period of $p \sim 1.7557$ days). This provides a critical accretion rate $\dot m_{c} \sim 4\times 10^{18}$ gm/s $\sim 0.3  \dot M_{Edd}$ \citep[Equ 5.106]{1986PhT....39j.124F}. If $\dot m < \dot m_{c}$ the source shows a transient behavior without any irradiation. Whereas for $\dot m > \dot m_{c}$, the irradiation forces the whole disc to be in the hot state and the disc finds a steady state with persistent activities \citep{1998MNRAS.293L..42K}. We see from Figure \ref{fig:Figure_15} that in the flat part $\dot m_d \gtrsim \dot m_{c}$, indicates quasi-steady activities after peak-I whereas 2004/05 outburst requires a lower value of $\dot m_c$. The corresponding sound speed is $c_{s} \sim 5.8\times 10^{5}$ cm/s at $R_{0}$ from the standard \citep{1973A&A....24..337S} picture. The hot front move inward with speed $\alpha_{h}c_{s}$, where  $\alpha_{h}$ is the viscosity parameter in the hot branch of the disc and the rising time of the outburst is $R_{0}/\alpha_{h}c_{s} \sim 88.6$ days (for $\alpha_{h}=0.1$). This is very similar to the observed result ($t_{3-6}$ and $t_{6-20}$ of Table \ref{lcpeak}) from multiple outbursts of the source.

Whereas the same scenario may not be applicable to another outbursting LMXB, GRO J1655-40 which has similar binary characteristics ($M\sim 7\textup{M}_{\odot}$, $q\sim 0.33$, $P \sim 2.62157$ days) as GX 339-4 but with much faster rising profile (rise time $\sim$ 12 days) for the primary outburst lightcurve. Such a short rise time indicates that outburst cannot be triggered at the outer edge of the Keplerian disc. Though \cite{2000A&A...354..987E} have applied irradiated DIM to explain the strongly flaring plateau state but without addressing the rise time. Alternately, an outburst may also initiate due to instability in the existing sub-Keplerian halo in quiescence by a sudden enhancement of viscosity which converts a fraction of the sub-Keplerian matter into Keplerian matter keeping the total accretion rate roughly constant \citep{1997ApJ...484..313C}. 
Using hydrodynamic simulation, \citep{2013MNRAS.430.2836G} have shown that sub-Keplerian matter can be converted into Keplerian matter with viscosity parameter above a certain critical value. Sub-Keplerian halo is generally optically thin and radiatively not very efficient, particularly in soft X-ray. That is why in LHS where the flow is mostly dominated by sub-Keplerian flow the source is low luminous. Whereas the same amount of matter in a Keplerian disc is radiatively more efficient because it could emit as a blackbody. Hence, during the evolution of outburst if sub-Keplerian matter is converting into Keplerian matter the overall luminosity increases. In addition, \cite{2010ApJ...710L.147M} have shown that the general HID profile of 2005 outburst of GRO J1655-40 can be explained using this proposal with an observed rise time of 12 days.
Also, they modeled the evolution of accretion rates as a powerlaw of time duration. But they did not actually fit the individual spectral data to justify the time evolution of accretion parameters. We model the time evolution of the accretion parameters and find GX 339-4 has a different behaviour than GRO J1655-40.

A logarithmic time response of accretion rates may be a signature of outburst with relatively longer rising time and very extended outer edge of the Keplerian disc. Hence, mass transfer at the outer edge continues along with the outburst. Whereas an  outburst with shorter rising time associates with powerlaw nature of mass transfer \citep{2015ApJ...804...87L, 2010ApJ...710L.147M} and having a smaller accretion disc. Also, in the later case the accumulated material at the outer part of the disc moves inward due to sudden viscous surge. We have tried to model the general profiles of parameters evolution not detailed day-to-day variations. In general, particularly in the intermediate states, the behaviour of the systems are highly non-linear and time dependent radiative-hydrodynamic simulations are required to address these issues properly. Nevertheless, the detail physics behind this process is unclear and in future we aim to apply this study on a few other outbursting sources to get a more clear picture.  

\subsection{A unified view of outbursts}
\label{unified}
One noticeable feature in the 3.0-20.0 keV light curves (Figure \ref{fig:Figure_1} is that 2002/03 and 2004/05 outbursts have two peaks whereas a second peak is not obvious in the other outbursts. We separately plot the 3.0-6.0 keV (soft) and 6.0-20.0 keV (hard) light curves for all outbursts in Figure \ref{fig:Figure_A1}. In all outbursts 6.0-20.0 keV light curves show two different peaks (I and II) whereas 3.0-6.0 light curves of 2002/03 and 2004/05 outbursts only show two peaks. The details of the light curves properties at peaks are summarized in Table \ref{lcpeak}. Another observable characteristic at peak-I is 3.0-6.0 keV light curve of every outbursts lags 6.0-20.0 keV light curve around 8-12 days. This can be understood from the different time scales controlling the soft and hard photons contributions which are discussed in \S\ref{accpara}. 
\begin{figure*}
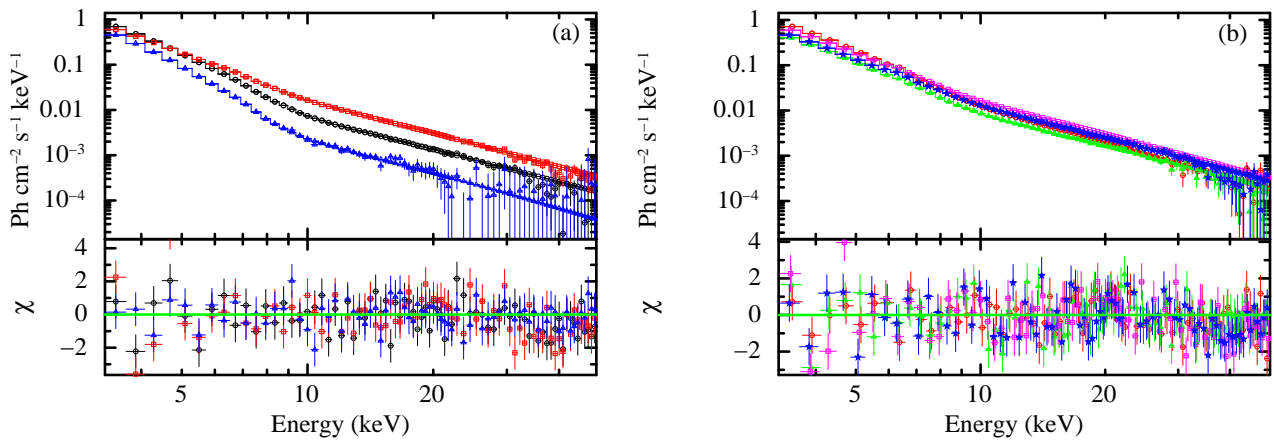

\includegraphics[width=0.33\textwidth,angle =-90]{./figure_16a.eps}
\includegraphics[width=0.33\textwidth,angle =-90]{./figure_16b.eps}
\caption{(a) Spectral fitting before (black-circle on MJD 54159.51), after (blue-triangle on MJD 54172.39) and at the peak-II (red-square on MJD 54161.67) of the hard flare during 2006/07 outburst. (b) Spectral characteristics at hard peak-II for 2002/03 (red-circle), 2004/05 (green-triangle), 2006/07 (magenta-square) and 2010/11 (blue-star) outbursts.}
\label{fig:Figure_16}
\end{figure*}

\begin{table}
\caption{Properties at peaks in light curves of different energy band}
\begin{tabular}{|l|l|l|l|l|l|}
\hline 
Outburst & Peak No & $F_{3-6} $ & $t_{3-6} $ & $F_{6-20}$ & $t_{6-20}$ \\ 
• & •  &  & (Day) &  & (Day)\\
\hline 
2002/03 & Peak I & 6.73 &  44 &  8.29 &  32\\  

 • & Peak II & 8.57 & 117 &  2.92 &  158\\ 

\hline 
2004/05 & Peak I & 2.29 & 95 &  2.49 & 88\\  
        & Peak II & 4.21  & 195 & 1.95 & 195\\ 

\hline 
2006/07 & Peak I & 8.71 & 94 & 8.00 & 86 \\  
        & Peak II  & --- & --- & 3.57 & 108 \\ 

\hline 
2010/11 & Peak I & 6.81 & 96 & 6.43 & 85 \\ 
        & Peak II & --- & --- & 2.79 & 110 \\  
\hline 
\end{tabular}
\ Note: $F_{3-6}$ and $F_{6-20}$ are fluxes in units of ${{10^{-9} erg/s/cm^{2}}}$ in 3.0-6.0 keV and 6.0-20.0 keV respectively. Whereas $t_{3-6}$ and $t_{6-20}$ are the time since triggering that the lightcurves peak in these two energy band respectively.
\label{lcpeak}
\end{table}
 
Unlike other outbursts of this source during the RXTE era, 2002 and 2004 outbursts are having very extended quasi-steady activities with a second peak (peak-II). The 2002 outburst witnessed a four year long quiescence phase in comparison with the outburst in every couple of years. This means the Keplerian disc got enough time to rebuild and extended much inside (smaller $Xs_0$ in Table \ref{fit_par}). Along with this a shorter rising time ($\sim 40$ days) and more symmetric rise-decay profile of the peak-I of 2002 outburst indicates that possibly the outburst did not trigger at the outer edge of the disc. Rather it started as an in-outward outburst which enhanced the mass accretion rate in Keplerian accretion disc and the disc has moved close to the central object faster, reaching the flux at peak-I. The irradiation of the outer disc due to central X-ray source keeps the outer disc hot and sub-Keplerian rate decrease sharply (Figure \ref{fig:Figure_14}a) due to enhanced viscosity at the outer edge. But after a few days, the sub-Keplerian rate halts to another constant value $\sim 0.03$ and $\dot m_h$ always remain constant during triggering of outburst. This indicates that the irradiation possibly affects the binary companion as well with an enhanced mass accretion at the outer part of the disc and hence sub-Keplerian rate does not decrease further. Considering that the second phase of mass transfer triggered after peak-I, the duration between the soft peak-II and hard peak-I is $\sim 85$ days which is comparable to the rising time of other outbursts. This continued accretion onto the central object produces a quasi-steady flaring activity till it reaches peak-II. Unlike the normal outburst triggering in LHS, the second triggering has happened in SIMS with a much higher value of $\dot m_d$ and a lower value of $\dot m_h$. After soft peak-II, $\dot m_h$ finally starts to decline again and finally the outburst starts dimming when almost the entire Keplerian disc mass is engulfed by the central BH. 
The 2004 outburst is different from 2002 outburst, having usual rise time $\sim 90$ days and faintest among all the outbursts. The 2004 outburst triggered at the outer edge of the disc and propagate as out-inwards due to the long rise time. The rest of the dynamics is very similar to 2002 outburst proposed above. The low luminosity of 2004 outburst is due to a lower Keplerian disc accretion (Figure \ref{fig:Figure_15}b) rate than other outbursts and possibly this indicates a lower disc mass prior to the outburst. Thus the existence of the second soft peak where the overall flux has a maximum in both 2002/03 and 2004/05 outburst, can be thought of a combination of two successive triggering of outbursts i.e., a second triggering of outburst may have happened after peak-I. The second triggering of 2002/03 and 2004/05 outburst are represented by a solid black line in Figure \ref{fig:Figure_14}a,b with the same parameters as in Table \ref{fit_par} but $t_h$ =150 days and 200 days respectively. Interestingly, \cite{1998MNRAS.293L..42K,1998MNRAS.301..382S}, suggested the presence of a secondary maxima after $\sim 10-15$ days of the overall peak in the outburst profile of BH XRBs due to irradiation in DIM. The thermal instability at the outer edge due to irradiation triggers small outburst, adding central accretion rate during decay phase. The 2006/07 and 2010/11 outbursts are more of classical irradiation DIM type with a hint of peak-II and here the sub-Keplerian matter continuously decrease (Figure \ref{fig:Figure_15}c,d) after the peak flux. Recently, \cite{2018MNRAS.480....2T,2018Natur.554...69T} have done notable works to unify the lightcurve profiles and have quantitatively estimated the $\alpha$-viscosity parameter for different XRBs by fitting the light curve with an irradiated accretion disc model. Though they did not apply their model for 2002 and 2004 outbursts of GX339-4 where double peaks have been observed.

The hard peak-II (Figure \ref{fig:Figure_A1}) have very interesting characteristics. We see that the hard flare (Peak-II) for all outbursts are appearing at a similar time with a time shifting for 2002 and 2004 outbursts. For 2006/07 and 2010/11 outbursts, they appear $\sim 108$ days after triggering. Since both 2002 and 2004 outbursts have two soft peaks corresponds two successive triggering discussed above, we consider the time zero to reach the hard peak-II corresponds to the time when $\dot m_h$ in Figure \ref{fig:Figure_14}a,b approach the second constant value (beginning of solid line). This is 50 days (2002 outburst) and 95 days (2004 outburst) after the first triggering respectively. Finally, in Figure \ref{fig:Figure_A1}e (middle panel), we plot the 6.0-20.0 keV hard flux after 50 days (2002) and 95 days (2004) onwards along with the other two outbursts. The red-solid, green-dot, magenta-dash and blue-dash dot lines represent the hard flares (peak-II) for 2002/03. 2004/05, 2006/07 and 2010/11 outburst respectively.  We see that Peak-II appears after $\sim$ 110 days in all cases i.e., the hard flaring signature after the maximum of soft flux in SIMS are generic to every outbursts of this source. This is possibly a direct indication of irradiation DIM origin of outburst for this source.
We study the spectral characteristic around the hard flare and as a representative case in Figure \ref{fig:Figure_16}a, we have shown PCA spectra before (black-circle), after (blue-triangle) and at peak-II (red-square) of 2006/07 outburst. It shows that the hard component of flux at peak-II is significantly more than the flux before and after the peak-II and we see a similar spectral behaviour around peak-II in all other outbursts. The existence of hard flares in SIMS \citep{2002MNRAS.331..765B,2003ApJ...592.1100T,2004ApJ...617.1272C,2005ApJ...629.1008J,2009A&A...501....1C,2018Ap&SS.363...90N} are observed in many other BH outbursting sources as well.
This may be due to the evacuation of the inner portion of the disc such that low energy flux reduces and the overall spectrum becomes harder. Also, radio jets are commonly observed \citep{2001MNRAS.322...31F,2002ApJ...573L..35C,2001MNRAS.322...31F} during hard flares in SIMS but no radio observations exist during hard peak-II of this source. In Figure \ref{fig:Figure_16}b, we have shown a comparison of PCA spectra during hard peak-II for all outbursts. The figure shows spectra with very similar spectral indices and flux values. Thus, all four outbursts can be unified with similar rising time (after introducing a second triggering for 2002/03 outburst), the appearance of hard flare at similar time scales even though they have very different flux at maxima.

\section{Conclusions}
\label{conclusions}
In this paper, we have studied both spectral and dynamical evolution of outbursting source GX 339-4. We have used RXTE as well as nearly simultaneous XRT/Swift and PN/XMM-Newton data during 2002-2011 for this study. The simultaneous broadband spectral modelling is shown in Figure \ref{fig:Figure_3}, \ref{fig:Figure_4}, \ref{fig:Figure_5}. The evolution of spectral parameters are presented in \S\ref{evo-para}. We see PCA data alone cannot constrain $T_{in}$ in LHS and simultaneous observations (red-dash points in Figure \ref{fig:Figure_6}d) towards lower energy are required. In all other states $T_{in}$ have similar values for all outbursts though {\it diskbb norm} (Figure \ref{fig:Figure_7}) have very different values. The powerlaw index (Figure \ref{fig:Figure_8}) evolve between $1.5-3$ during LHS to HSS. It remains constant in LHS and follow an anti-clockwise hysteresis with the 6.0-20.0 keV flux. The {\it powerlaw norm} (Figure \ref{fig:Figure_9}) shows high values during the rising phase and hard flare in SIMS. In Figure \ref{fig:Figure_10}, we observe that the Fe line flux is week in LHS and the strength increase towards softer states. The line normalization (Figure \ref{fig:Figure_11}) shows a decreasing requirement of line during successive outbursts after 2002/03 and the minimum value of the line normalization increases for later outbursts.

The spectral modeling using two-component accretion flow is presented in \S\ref{simu-phy}. We can understand the evolution of the spectral parameters (\S\ref{evo-para}) from the evolution of the accretion parameters (\S\ref{accpara}). Figure \ref{fig:Figure_15} and Figure \ref{fig:Figure_6} show that as $\dot{m_{d}}$ increases $T_{in}$ also increases in the rising phase and both decrease in the declining phase. The photon index (Figure \ref{fig:Figure_8}) remains constant in LHS as $\dot{m_{h}}$ is constant (Figure \ref{fig:Figure_14}) and $\dot{m_{d}}$ has a low value (Figure \ref{fig:Figure_15}). As $\dot m_d$ increases, the supply of soft photons becomes more and it cools the POSR which decreases the 6.0-20.0 keV flux and the spectral index increase towards softer states. The reflection signature (and hence line flux) is low (Figure \ref{fig:Figure_10}) in LHS as the source is faint and the Keplerian disc is far apart. As the source evolves towards softer state, the Keplerian disc reaches closer to the source. Also the presence of significant hard flux from the POSR makes the line flux strongest in SIMS. We can understand the dynamical evolution of the system during all outbursts from Figures \ref{fig:Figure_13}, \ref{fig:Figure_14}, \ref{fig:Figure_15}. We present a logarithmic evolution of accretion rate (Equation \eqref{eq:md} and  Equation \eqref{eq:mh}) to understand the outburst evolution and expect to hold for outburst with longer duration. In this case, the total accretion rate is not constant and Keplerian disc extends to very large distance. We have explained the evolution of all the outburst of GX 339-4 in the light of irradiate DIM using two-component accretion flow (\S\ref{accpara}). For outbursts with shorter duration may have a smaller accretion disc and the accreting matter pileup at the outer edge. The sudden enhancement of viscosity triggers the outbursts with a conversion of sub-Keplerian matter into Keplerian matter and keeping the total accretion rate roughly constant. In this paper, we have tried to understand the general behaviour of accretion parameters evolution and the detailed small scale variations are outside the scope of the paper. In Figure \ref{fig:Figure_13}, \ref{fig:Figure_14}, \ref{fig:Figure_15}, we have done a manual fitting to understand the behaviour of the general profile and did not estimate the errors in the parameters given in Table \ref{fit_par}. This is one of the caveats in the current study.

The observed light curves of 2002/03 and 2004/05 outbursts appear very different from 2006/07 and 2010/11 outbursts. But introducing a second triggering of outburst after the hard peak-I due to irradiation from the central X-ray source resembles that soft peak-II in 2002/03 and 2004/05 outbursts appear at a similar rising time of 2006/07 and 2010/11 outbursts. This unified picture of all four outbursts are supported by the two-component model fitting as well (Figure \ref{fig:Figure_14}a,b) by halting the decreasing trend of $\dot m_h$ after peak-I. The presence of hard flares in SIMS have been observed in many outbursting galactic black hole sources. But the appearance of hard flares (Figure \ref{fig:Figure_A1}e) around similar time ($\sim 110$ days) after triggering (from second triggering time for 2002 and 2004 outbursts) for all outbursts is associated with the intrinsic nature of this source that the soft flux reaches the maxima at similar time scale. This can be attributed to the irradiation in the accretion disc by the central source.
In future, we would like to implement these two pictures in other outbursting sources as well and may be able to get a better understanding of these systems. 

\section*{Acknowledgements}
We thank the anonymous referee for giving valuable comments that have significantly improved the quality of the paper. 

\bibliography{refer}
\appendix
\section*{Appendix}
\renewcommand{\thetable}{A\arabic{table}}
\begin{table}
\caption{XMM-Newton, XRT/Swift and RXTE simultaneous observation catalog during 2002-2011}
\resizebox{\columnwidth}{!}{%
\begin{tabular}{|c|c|c|c|}
\hline
&\multicolumn{2}{|c|}{2002/03 outburst}&\\
\hline
XMM Obs Id  & XMM Obs date (MJD) & RXTE Obs Id & RXTE Obs date (MJD) \\
\hline
0148220301    &    52718.055  &  50117-01-04-02  &   52718.173 \\
0148220201    &    52706.731  &  50117-01-03-01  &   52706.840  \\
\hline \hline
&\multicolumn{2}{|c|}{2006/07 outburst}&\\
\hline
XRT Obs Id    &  XRT Obs date (MJD)  &  RXTE Obs Id   &    RXTE Obs date (MJD) \\
\hline
00030919001 	&54213.031 &	92704-03-02-00 &	54213.387 \\
00030919002	&     54216.520 &	92704-03-01-02 &	54216.939 \\
00030919003	&     54220.071 &	92704-03-05-01 &	54220.226 \\
00030919005 	&	54227.298 &	92704-03-08-01 &	54227.613 \\
00030919006	&     54231.913 &	92704-03-10-00 &	54231.604 \\
00030919007	&     54234.922 &	92704-03-11-00 &	54234.837 \\
00030919008	&     54237.331 &	92704-04-01-03 &	54237.290 \\
00030919010	&     54240.884 &	92704-03-13-02 &	54240.301 \\
00030919012	&     54244.715 &	92704-03-14-03 &	54244.971 \\
00030943001	&     54245.711 &	92704-04-02-00 &	54245.749 \\
\hline \hline
&\multicolumn{2}{|c|}{2010/11 outburst}&\\
\hline
XRT Obs Id    &  XRT Obs date (MJD)  &  RXTE Obs Id   &   RXTE Obs date (MJD) \\
\hline
00030943005 	& 55217.673	& 95409-01-02-02 &	55217.810 \\
00030943006	& 55259.964	& 95409-01-08-03 &	55259.791 \\
00030943007	& 55260.240	& 95409-01-09-04 &	55260.071 \\
00030943008	& 55260.518	& 95409-01-09-05 &	55260.445 \\
00030943010	& 55261.044	& 95409-01-09-06 &	55261.189 \\
00030943011	& 55281.744	& 95409-01-12-00 &	55281.581 \\
00030943012	& 55285.733	& 95409-01-12-02 &	55285.778 \\
00030943013	& 55289.302	& 95409-01-13-00 &	55289.618 \\
00030943014	& 55293.036	& 95409-01-13-01 &	55293.087 \\ 
00030943015	& 55297.782	& 95409-01-14-02 &	55297.874 \\
00030943016	& 55301.388	& 95409-01-14-07 &	55300.919 \\
00030943017	& 55301.789	& 95409-01-14-05 &	55301.790 \\
00030943018 	& 55302.411	& 95409-01-15-00 &	55302.196 \\ 
00030943019	& 55302.800	& 95409-01-15-00 &	55302.196 \\
00030943020	& 55303.809	& 95409-01-15-01 &	55303.604 \\
00031687002	& 55316.986	& 95409-01-17-00 &	55316.115 \\
00031687005	& 55334.061	& 95409-01-19-04 &	55334.382 \\
00031687010	& 55366.381	& 95409-01-24-01 &	55366.514 \\
\hline
\end{tabular}
}
\label{xmm_xrt_pca_obs}
\end{table}

\renewcommand{\thefigure}{A\arabic{figure}}

\begin{figure*}
\includegraphics[width=1.13\textwidth]{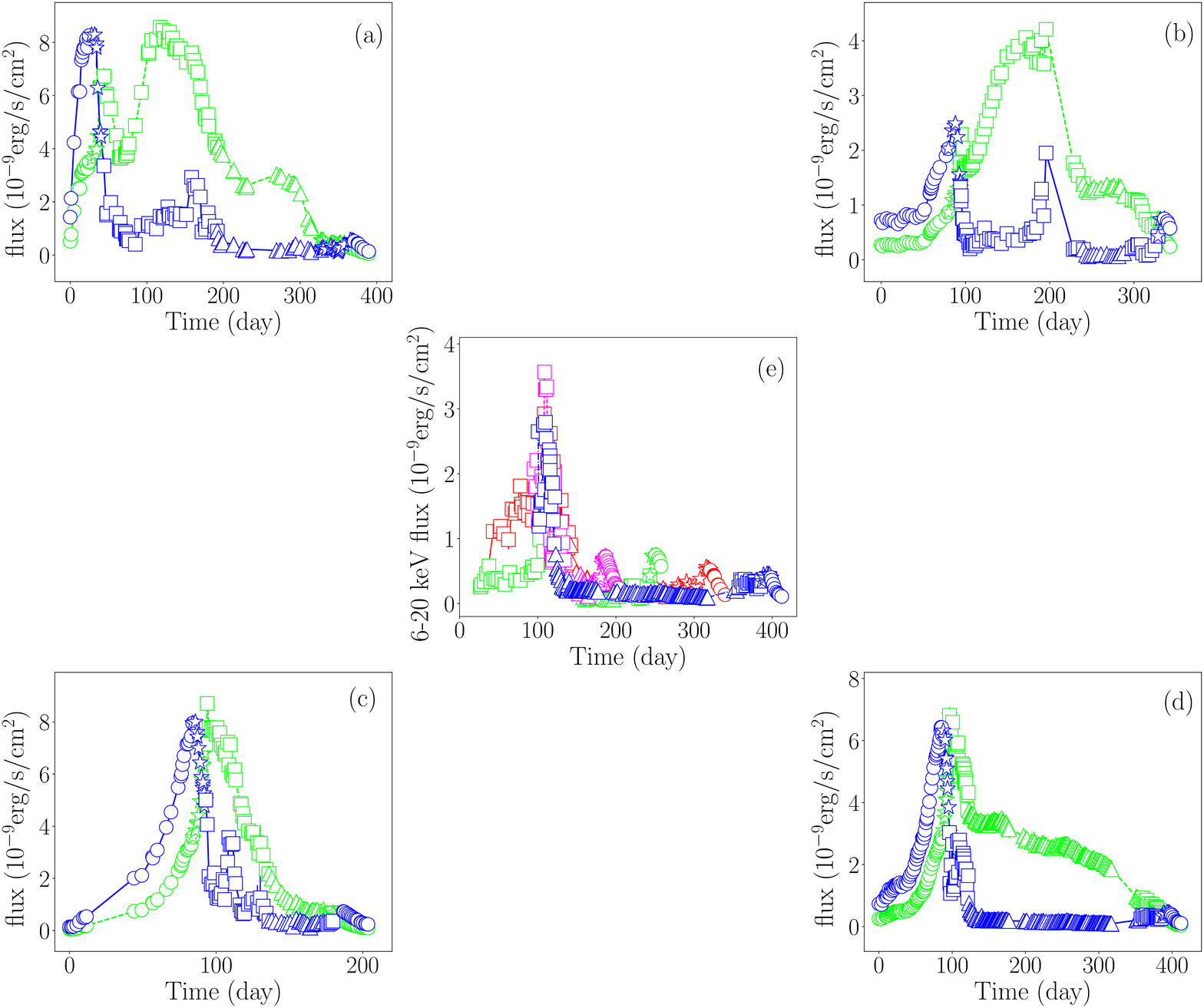} 
\caption{3.0-6.0 keV (green-dash) and 6.0-20.0 keV (blue-solid) PCA light curves of GX 339-4 during (a) 2002/03, (b) 2004/05, (c) 2006/07 and (d) 2010/11 outburst respectively. The different symbols: circle marks LHS; star marks HIMS; square marks SIMS and triangle marks HSS. Panel (e) shows the evolution of 6.0-20.0 keV hard flux after 50 days (red-solid:2002) and 95 days (green-dot:2004) onwards along with other two outbursts. In that case, the hard peak-I for 2006/07 and 2010/11 outbursts also should appear in the figure but it makes the figure very clumsy. Hence for presentation purpose, we have plotted 6.0-20.0 keV flux for 2006/07 (magenta-dash) and 2010/11 (blue-dash dot) outbursts after hard peak-I with time zero at the beginning of the outburst.}
\label{fig:Figure_A1}
\end{figure*}
\end{document}